\documentclass[]{agujournal2018}
\usepackage{apacite}
\usepackage{url} 
\usepackage[T1]{fontenc}
\usepackage[latin9]{inputenc}
\usepackage[english]{babel}
\usepackage{float}
\usepackage{textcomp}
\usepackage{amsmath}
\usepackage{graphicx}
\usepackage{esint}
\usepackage{natbib}

\draftfalse

\journalname{Geophysical Research Letters}

\begin{document}

%
%


\title{\textcolor{black}{Comparing} Convective Self-Aggregation in Idealized Models to Observed
Moist Static Energy Variability near the Equator}

%
%




\authors{Tom Beucler \affil{1,2}, Tristan H. Abbott \affil{3}, Timothy W. Cronin \affil{3}, Michael S. Pritchard \affil{1}}

 \affiliation{1}{Department of Earth System Science, University of California, Irvine, CA, USA}
 \affiliation{2}{Department of Earth and Environmental Engineering, Columbia University, New York, NY, USA}
 \affiliation{3}{Department of Earth, Atmospheric and Planetary Sciences, MIT, Cambridge, MA, USA}





\correspondingauthor{Tom Beucler}{tom.beucler@gmail.com}


\begin{keypoints}
\item Moist static energy zonal spectral tendencies have similar signs and scale-selectivity in \textcolor{black}{convection-permitting} models and observations.
\item Radiation increases variance at long wavelengths while surface enthalpy fluxes and advection reduce variance.
\end{keypoints}



%
%


\begin{abstract}
Idealized \textcolor{black}{convection-permitting} simulations of radiative-convective equilibrium (RCE) have become a popular tool for understanding the physical processes leading to horizontal variability of tropical water vapor and rainfall. However, the applicability of idealized simulations to nature is still unclear given that important processes are typically neglected, such as lateral vapor advection by extratropical intrusions, or interactive ocean coupling. Here, we exploit \textcolor{black}{spectral analysis to compactly summarize} the multi-scale processes supporting convective aggregation. By applying this framework to high-resolution reanalysis data and satellite observations in addition to idealized simulations, we compare convective-aggregation processes across horizontal scales and datasets. The results affirm the validity of the RCE simulations as an analogy to the real world. Column moist static energy tendencies share similar signs and scale-selectivity in \textcolor{black}{convection-permitting} models and observations: Radiation increases variance at wavelengths above 1,000km, while advection damps variance across wavelengths, and surface fluxes mostly reduce variance between 1,000km and 10,000km. 
\end{abstract}

%
%
\section*{Plain Language Summary}

Advances in computing have allowed computer models to simulate tropical weather systems spanning a few dozen kilometers at the same time as moist and dry regions spanning several thousand kilometers. To improve and validate computer models, we need to compare computer simulations to real-world observations, but we lack a compact way of simultaneously comparing them at scales close to 10km, 100km, 1,000km, and 10,000km. By breaking down water vapor variability near the Equator into contributions from these different length scales, we can identify the scales at which computer models agree with real-world observations and explain why. Surprisingly, even computer models that are run in a highly idealized configuration \textcolor{black}{compare} well against observations of the real world, despite the fact that nature never attains this idealized limit. We find that atmospheric radiation tends to intensify moist and dry regions of several thousand kilometers near the Equator, while lateral transport of energy and surface-atmosphere exchanges tend to smooth out these moist and dry regions. 

\section{Introduction}

\begin{figure}[H]
\begin{centering}
\includegraphics[width=15cm]{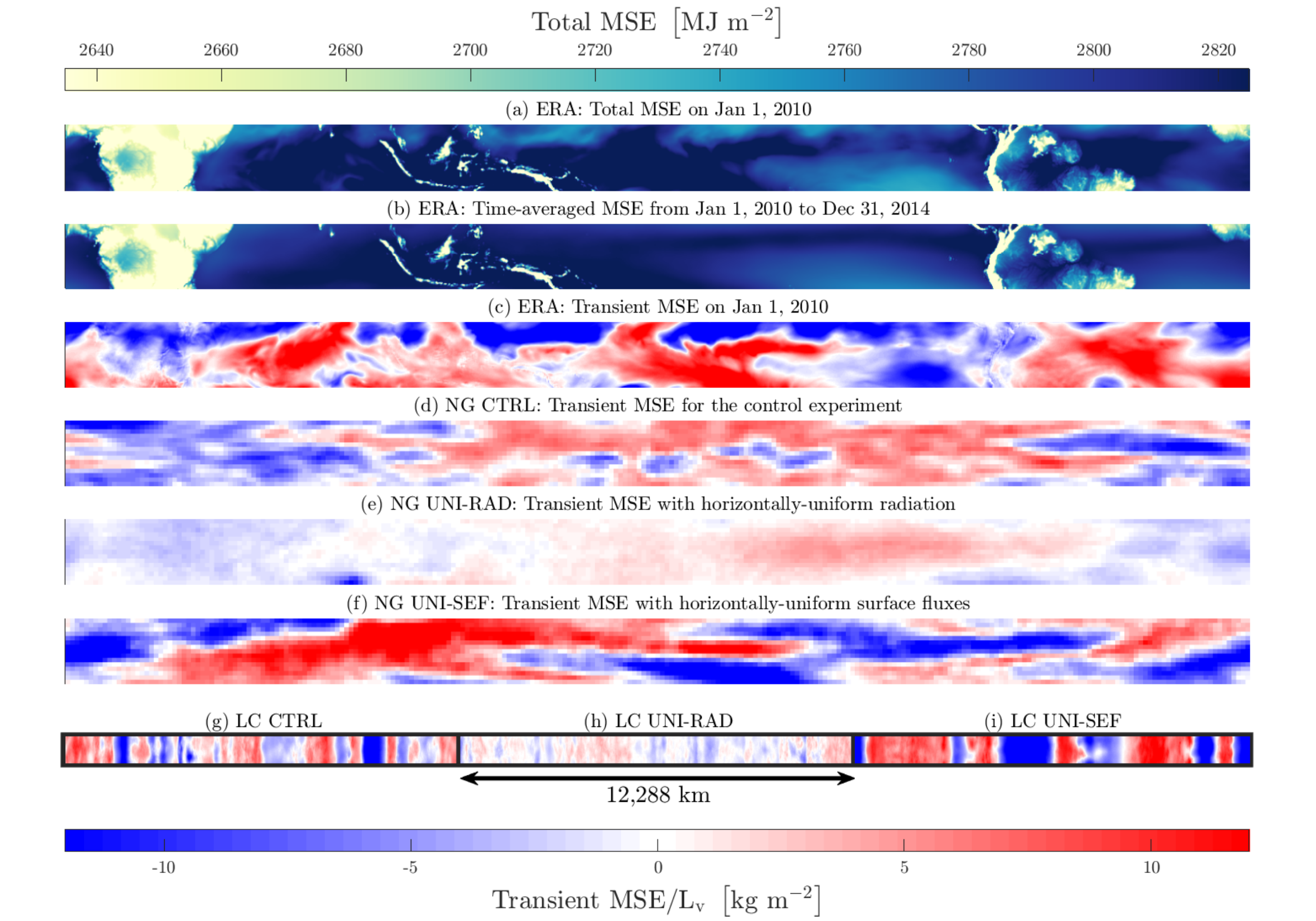}
\par\end{centering}
\caption{(a) Instantaneous, (b) time-averaged, and (c) transient MSE in ERA from $10\text{\textdegree}$S to $10\text{\textdegree}$N. (d-i) Instantaneous transient MSE in each model of section \ref{sec:Data}. Transient MSE is normalized by the latent heat of vaporization of water $L_{v}\ $to yield units $\textnormal{kg}\ \textnormal{m}^{-2}\ $: the length of the bottom colorbar corresponds to $\sim60\textnormal{MJ\ m}^{-2}$. Panels (a-f) respect the original aspect ratio of the horizontal domain. While the length of the long-channel equals a third of the Equator's length, its width has been multiplied by a factor $5\ $in panels (g-i) to facilitate visualization. \label{fig:Snapshots}}
\end{figure}

Tropical weather and climate are strongly shaped by the variability of column water vapor, which dominates column-integrated moist static energy (MSE) variability due to weak horizontal variations in tropical atmospheric temperature. On meteorological timescales, the intensity of extreme precipitation events depends on the humidity and temperature of the surrounding environment, e.g. for isolated convective cells, mesoscale convective systems \citep{LeMone1998} and tropical cyclones \citep{Hill2009}. On climatic timescales, the zonal variability of MSE is linked to the equator-to-pole energy transport \citep{Trenberth2002} and to climate sensitivity through the link between the hydrological cycle and cloud and water vapor feedbacks \citep{Feldl2014}. Persistent regions of high and low MSE occur due to surface heterogeneities, including ocean currents, continents, and mountain ranges (Figure \ref{fig:Snapshots}a), while transient anomalies in MSE near the Equator (e.g., Figure \ref{fig:Snapshots}c) relate to a rich spectrum of tropical weather across a range of temporal and spatial scales. This includes isolated convective activity ($\sim$1 hour, $\sim10$ km), mesoscale convective complexes ($\sim$10 hours, $\sim100$ km) \citep[e.g. review by][]{Houze2004}, tropical depressions ($\sim$ 10 days, $\sim1000$ km) \citep[e.g. review by][]{Montgomery2017}, \textcolor{black}{the Madden-Julian Oscillation \citep[e.g. review by][]{Zhang2005}, and the Asian monsoon ($\sim$ 60 days, $\sim10000$ km) \citep[e.g. review by][]{Webster1998}}. 

Advanced computing now allows simulation of planetary-scale domains ($\sim10^4$ km) with \textcolor{black}{convection-permitting} models (CPM) of horizontal resolution $\sim 1$ km, which can resolve this entire spectrum of tropical weather. \textcolor{black}{Tropical weather systems have been extensively compared in field-campaign observations and regional CPM to evaluate CPMs' ability to adequately represent convective processes \citep[e.g., ][]{Beucher2014,Laing2012,Woodhams2018}. However, explicit comparisons of physical processes regulating the spatio-temporal spectrum of MSE} in observations and CPM are rare. The goal of this paper is to use a spectral budget for sources and sinks of transient MSE variance as a step towards comparing \textcolor{black}{these physical processes} across observations and models of varying complexity.

We use the column-integrated frozen moist static energy $H$ (units J m$^{-2}$)\textcolor{black}{:  
\begin{linenomath*}
\begin{equation}
H\left(x,y,t\right)\overset{\mathrm{def}}{=}\int_{0}^{p_{s}}\frac{dp}{g}\left(L_{v}q-L_{f}q_{i}+\underbrace{c_{p}T+gz}_{s}\right),
\end{equation}
\end{linenomath*}
}as a diagnostic because it is \textcolor{black}{approximately conserved during convection}, and because previous studies in idealized CPM have successfully used its variance budget to assess processes that favor or disfavor convective aggregation \textcolor{black}{ \citep[e.g., ][]{Wing2014,Wing2016a}}. Here, \textcolor{black}{$ p_{s}$ is surface pressure and $ p$ the mean pressure profile}, $L_{v}$ and $L_f$ are the latent heat of vaporization and fusion of water, respectively, $q$ and $q_i$ are water vapor and ice mixing ratios, respectively, $c_{p}$ is the specific heat capacity of dry air at constant pressure, $T$ is the absolute temperature, \textcolor{black}{$g $ is the gravity constant, $z $ is the geopotential height, }and $s$ is the dry static energy. The total MSE field $H$ has spatial variability in its temporal mean $\overline{H}$, as well as spatiotemporal variability in the transient MSE anomaly $H^{\prime}$, here defined by: 
\begin{linenomath*}
\begin{equation}
H\left(x,y,t\right)=\overline{H}\left(x,y\right)+H^{\prime}\left(x,y,t\right)\label{eq:Transient_definition}
\end{equation}
\end{linenomath*}
Note that transient MSE variability may be modulated nonlinearly by the stationary MSE features, adding another level of complexity to MSE transients --- but in this paper we will focus primarily on comparing transient MSE variability across models and observations without directly assessing this role of nonlinear modulation. 

Previous work \citep[e.g.,][]{Held1993,Muller2015a} has consistently found that when CPMs are run on large enough domains, MSE self-organizes into moist and dry regions even in the absence of external forcing \textcolor{black}{(such as planetary rotation, surface inhomogeneities or large-scale wind shear)}. This emergent property of moist convection, referred to as ``convective self-aggregation'' \citep[e.g., review by][]{Wing2017,Holloway2017}, suggests that a significant fraction of transient MSE variability near the Equator might arise from internal self-organization rather than external processes such as surface characteristics, teleconnections with the mid-latitudes, or ocean coupling. The problem is that physical mechanisms of convective self-aggregation have been extensively studied in the context of idealized CPM with fixed surface temperatures, which ignore external processes, and thus provide an uncertain analogy to real-world settings. 

This motivates the aim of our paper --- quantitatively comparing convective-aggregation processes in idealized CPM and observations\textcolor{black}{. This comparison} may deepen our understanding of (1) how transient MSE anomalies grow and decay and (2) how valid \textcolor{black}{idealized CPM simulations are as an analogy to the real world}. Idealized CPM have been compared to observations in the past, but mostly at coarse granularity by looking for similar correlations or distributions of variables.
Using satellite data, \citet{Tobin2012a} showed that $\left(10^{\circ}\times10^{\circ}\right)\ $longitude-latitude boxes with more convective organization also exhibited lower values of MSE and larger outgoing longwave radiation, consistent with idealized CPM experiments \citep{Wing2014}. \citet{Holloway2017} used data from the Nauru meteorological station and showed that the long-channel configuration of \citet{Wing2016a} had more realistic  distributions of MSE and vertical velocity than traditional square-domain CPM. Additionally, \citet{Stein2017} showed that for a given large-scale precipitation rate and vertical motion, anvil clouds decreased with the degree of aggregation in satellite data, while low clouds and precipitation efficiency increased with aggregation, consistent with CPM simulations \citep[e.g., Figure 8 of][]{Wing2016a}. Recently, \citet{Holloway2017a} used the MSE spatial variance budget to show that interactive radiation maintained aggregation while MSE advection disaggregated convection in \textcolor{black}{CPM experiments forced using satellite data, as found in idealized CPM simulations of RCE}. 

\textcolor{black}{While Holloway's simulations support the validity of the RCE analogy, their small domain size $\left(10^{\circ}\times10^{\circ}\right)\ $ may underestimate convective-aggregation feedbacks because of the effect of MSE advection from the prescribed boundary conditions.} Motivated by the recent availability of planetary-domain CPM and high-resolution reanalysis products, we proceed by comparing the observed transient MSE field (Figure \ref{fig:Snapshots}c) to the transient MSE field from several idealized\textcolor{black}{, large-domain} CPM experiments\textcolor{black}{ \citep[][ Figures \ref{fig:Snapshots}d-i]{Wing2017,Khairoutdinov2018}} and ask:

How do the physical \textcolor{black}{processes} that \textcolor{black}{regulate} observed moist static energy variance compare to the convective-aggregation \textcolor{black}{processes} from idealized models \textit{at each horizontal scale}?

The work below is organized as follows. After introducing the observational and model datasets in section \ref{sec:Data}, we investigate the zonal power spectra of transient MSE and how they evolve under the influence of radiation, surface enthalpy fluxes, and advection in section \ref{sec:Zonal-Spectral-Budget-MSE}, before 
concluding in section \ref{sec:Conclusion}.

\section{Data\label{sec:Data}}

We use four datasets to compare convective aggregation in observations and idealized CPM: Meteorological reanalysis (ERA), satellite observations (CERES), a rotating near-global simulation (NG) and a non-rotating long-channel simulation (LC). A snapshot of the transient MSE field from each is shown in Figure \ref{fig:Snapshots}, and each is described in more detail below.





\subsection{Reanalysis observations: ERA}

The European Centre for Medium-Range Weather Forecasts Re-Analysis (ERA) version 5 \citep{Hersbach2016} was produced by assimilating observational data in version CY41R2 of the Integrated Forecast System. The new reanalysis dataset has a better hydrological cycle and sea surface temperatures in the Tropics and is calibrated for climate applications. Zonal and temporal resolutions are $27.5\textnormal{km}\times1\textnormal{hour}$. 

\subsection{Satellite observations: CERES}

The Clouds \& Earth's Radiant Energy Systems \citep[CERES, ][]{Wielicki1996} ``CERES SYN1deg Ed4A'' dataset provides diurnally-complete top-of-atmosphere and surface radiative fluxes by using sixteen geostationary satellites as well as the National Aeronautics and Space Administration's Moderate Resolution Imaging Spectroradiometer. Zonal and temporal resolutions are \textcolor{black}{$1^{\circ}\times1\textnormal{hour}$}. 

\subsection{\textcolor{black}{Convection-permitting} Model (CPM) Simulations}
The following experiments were conducted using the System for Atmospheric Modeling (SAM), a \textcolor{black}{convection-permitting} model that is widely used for idealized studies, solves the anelastic equations of motion, and includes cloud microphysics and subgrid turbulence parameterizations \citep{Khairoutdinov2003}. 
\begin{itemize}
\item \textbf{LC: } A suite of three idealized long-channel (LC) experiments with \textcolor{black}{ a doubly-periodic} horizontal domain \textcolor{black}{of }size $12,288\times192\textnormal{km}^{2}$ over a uniform ocean surface with temperature 300K, using SAM v6.8.2. These consist of a control simulation (LC CTRL, Figure \ref{fig:Snapshots}g), described in \citet{Wing2016a}, \textcolor{black}{and simulations that horizontally homogenize either radiation (LC UNI-RAD, Figure \ref{fig:Snapshots}h), or surface enthalpy fluxes (LC UNI-SEF, Figure \ref{fig:Snapshots}i), described in \citet{Beucler2018d}}. Each experiment was run for 80 days to a statistically steady state and outputs were saved with zonal and temporal resolutions of $3\textnormal{km}\times1\textnormal{hour}$.  
    \item \textbf{NG: } Similar to LC, but for a much larger (near-global; NG, $40,360\times10,000\textnormal{km}^{2}$) ocean-only\textcolor{black}{, zonally-periodic }domain with prescribed ocean surface temperatures that decrease away from the equator and a Coriolis parameter that increases away from the equator, allowing for formation of extratropical eddies which intrude into the tropics. As with LC, three runs were conducted at 300K, consisting of NG CTRL (control experiment, Figure \ref{fig:Snapshots}d), NG UNI-RAD (horizontally-uniform radiative heating, Figure \ref{fig:Snapshots}e) and NG UNI-SEF (horizontally-uniform surface enthalpy fluxes, Figure \ref{fig:Snapshots}f) in \citet{Khairoutdinov2018}. Each experiment was run for a year, using SAM v6.10.6. Outputs were saved with zonal and temporal resolutions of $156.25\textnormal{km}\times1\textnormal{day}$. 

\end{itemize}

At spatial scales below $\mathrm{O\left(100km\right)}$, most of the spatial variance comes from sub-diurnal variability from isolated convective events (not shown). Hence, to make meaningful comparisons across datasets, we time-average the fields of ERA, CERES and LC over one-day blocks before calculating spatial co-spectra using the Fast Fourier Transform algorithm \citep{Frigo2005}.

\section{Zonal Spectral Budget of Transient Column Moist Static Energy\label{sec:Zonal-Spectral-Budget-MSE}}

The following spectral method will allow us to (1) \textcolor{black}{separate the zonal variability of MSE in each dataset into contributions from different scales} and (2) quantify the amount of variance created by radiation, surface enthalpy fluxes, and advection at each zonal scale. Specifically, we measure zonal variability of transient MSE $H^{\prime}\ $at a given zonal wavelength $\lambda\ $using the zonal power spectrum $\varphi_{H}\ $of transient MSE, defined as: 
\begin{linenomath*}
\begin{equation}
\varphi_{H}\left(\lambda,y,t\right)\overset{\mathrm{def}}{=}\widehat{H^{\prime}}^{*}\widehat{H^{\prime}},
\end{equation}
\end{linenomath*}
where $t\ $is time and $\widehat{H^{\prime}}\ $ is the zonal Fourier transform of the transient MSE field $H^{\prime}$:
\begin{linenomath*}
\begin{equation}
\widehat{H^{\prime}}\left(\lambda,y,t\right)\overset{\mathrm{def}}{=}\frac{1}{\sqrt{2\pi}}\int_{0}^{L\left(y\right)}\exp\left(-\frac{2\pi\imath x}{\lambda}\right)H^{\prime}\left(x,y,t\right)dx,\label{eq:Fourier_transform}
\end{equation}
\end{linenomath*}
where $\imath\ $is the unit imaginary number and $L\left(y\right)\ $ is the length of the latitude circles of ordinate $y$ \textcolor{black}{in all cases but LC, for which $L\left(y\right)\ $ is the periodic domain's length}. From $\varphi_{H}$, one can calculate various aspects of the transient MSE zonal variability, including its spectral-mean wavelength (equation 2 of \citet{Beucler2018d}) and its total (i.e. wavenumber-integrated) zonal variance. 

\subsection{Zonal Power Spectra}

\begin{figure}[H]
\begin{centering}
\includegraphics[width=13cm]{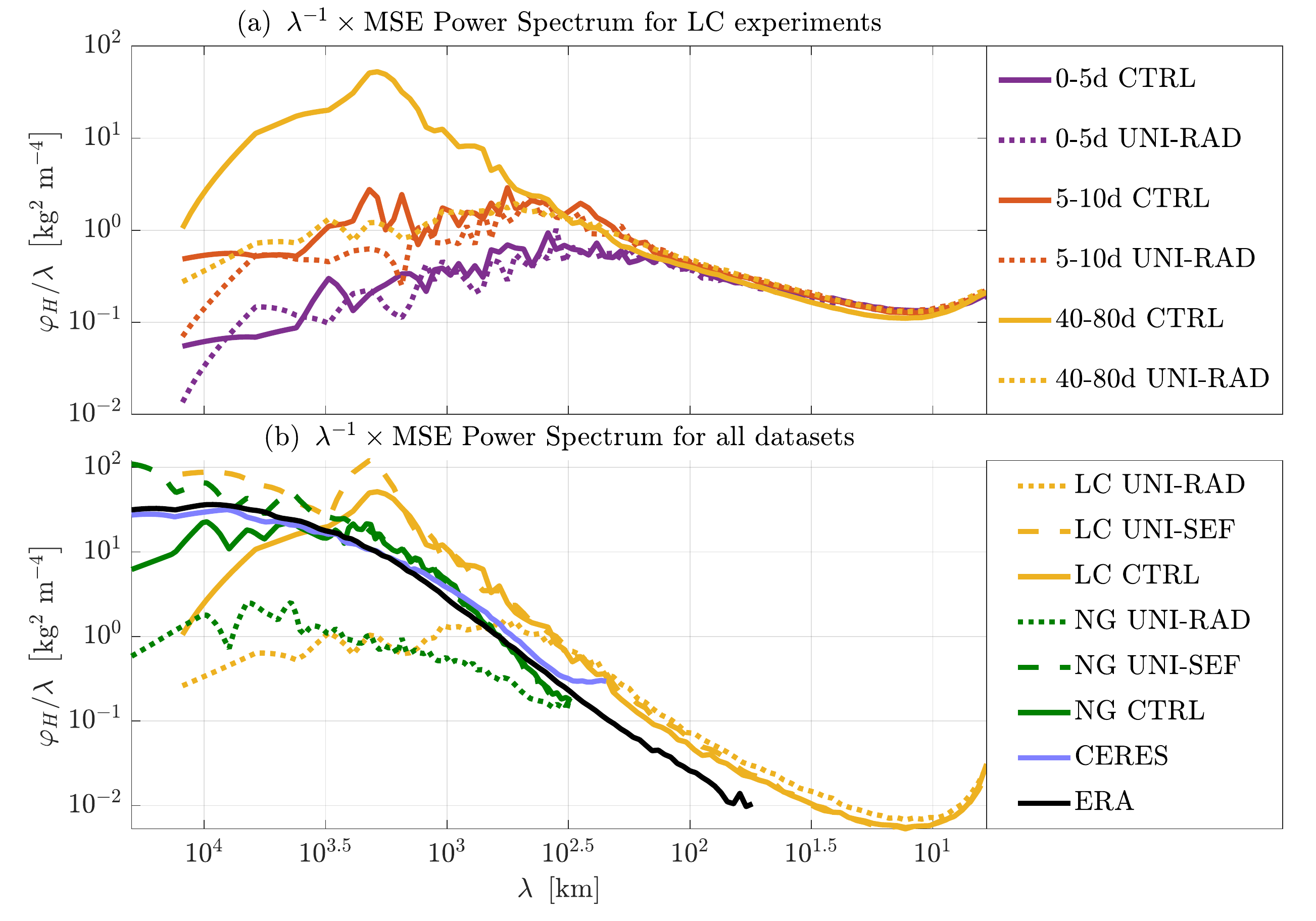}
\par\end{centering}

\caption{(a) Zonal power spectrum of transient MSE in the LC CTRL experiment (full lines) and the LC UNI-RAD experiment (dotted lines), time-averaged over different stages of the simulation. (b) Zonal power spectrum of transient MSE of all datasets, time-averaged over \textcolor{black}{40d}-80d for the LC experiments and over the entire time period for all other experiments. In both panels, spectra have been averaged over the $y-$dimension, and divided by $\lambda$ so that they integrate to the total variance in logarithmic $\lambda-$space. \textcolor{black}{To facilitate interpretation, spectra are divided by $L_{v}^{2} $ to yield units $\textnormal{kg}^2\ \textnormal{m}^{-4}$. Note that the observed spectra (light-blue and black lines) resemble the aggregated idealized spectra (yellow and green solid lines), but differ from the non-aggregated spectra (yellow and green dotted lines) by more than an order of magnitude at long wavelengths}. \label{fig:MSE_spectrum}}
\end{figure}

Figure \ref{fig:MSE_spectrum}a shows $\varphi_{H}\ $for two LC experiments: 
\begin{enumerate}
\item The control experiment CTRL (full lines): This experiment is initialized with a horizontally uniform sounding taken from small-domain RCE, but moist and dry regions of finite size ($\sim2,000\textnormal{km}$) and MSE anomalies ($\sim7\textnormal{kg\ m}^{-2} \times L_{v}$) spontaneously form after $\sim1$month despite homogeneous boundary conditions \footnote{\textcolor{black}{See Figure \ref{fig:Snapshots}g for a snapshot at t=1month and Figure 1d of \citet{Beucler2018d} for a Hovmoller plot of the full time-evolution}}. Although the temporal variations of the transient MSE field appear complicated in physical space, Figure \ref{fig:MSE_spectrum}a reveals a simpler picture in spectral space: As convection self-aggregates (i.e. progressing from solid purple to yellow lines \footnote{\textcolor{black}{See Figure 4b of \citet{Beucler2018d} for the time-evolution of the total MSE variance}}), MSE variance increases at wavelengths above $\lambda\sim100\textnormal{km}$ before equilibrating with a variance peak at $\lambda\sim2,000\textnormal{km}$, explaining why anomalies of this scale are most visible in Figure \ref{fig:Snapshots}g. Note that the y-axis of Figure \ref{fig:MSE_spectrum}a is logarithmic so the total variance in the aggregated state (solid yellow line) is dominated by the $\lambda\sim2,000\textnormal{km}\ $variance peak.
\item The UNI-RAD experiment (dotted lines): Horizontally homogenizing radiative heating greatly weakens aggregation, as evidenced by reduced MSE perturbations ($\sim2\textnormal{kg\ m}^{-2}\times L_{v}$). Unlike the CTRL experiment, MSE variance only grows at the longest wavelengths for the first $10 $ days before stabilizing around $1-2\textnormal{kg}^{2}\ \textnormal{m}^{-4}\times L_{v}^{2}$. Using UNI-RAD as our reference ``non-aggregated'' experiment\footnote{This experiment is very similar to the experiment in which both radiative heating and surface enthalpy fluxes are horizontally uniform (see Figures 1a and 1c of \citet{Beucler2018d})}, the effect of self-aggregation can then be quantified as the difference between the full and dotted yellow lines, and is only significant on scales larger than $\lambda\sim1,000\textnormal{km}$.
\end{enumerate}
Moving to the more realistic NG simulations in Figure \ref{fig:MSE_spectrum}b, the effect of self-aggregation is qualitatively similar to the LC case if measured by the difference between the full and dotted green lines. That is, removing the spatial variability of radiation (green dotted line) prevents the transient MSE field from developing variance at long wavelengths relative to the control (green solid line). In contrast, removing the spatial variability of surface enthalpy fluxes adds variance at long wavelengths (dashed lines) in both the LC and NG setups, because surface enthalpy fluxes damp developing MSE anomalies after the initial stages of aggregation, opposite to radiation \citep[see ][ for an extensive discussion on this topic]{Beucler2018d}. 

We now turn to our main goal of comparing idealized simulations against observational data. The black line in Figure \ref{fig:MSE_spectrum} depicts the observational ERA spectrum, averaged from 10\textdegree S to 10\textdegree N and over 5 full years (January 1st 2010 -- December 31st 2014). The zonal MSE variance is slightly larger in ERA than in the NG CTRL case, except over the range of wavelengths where the NG spectrum peaks $\left(\lambda\sim500\textnormal{km}-2,500\textnormal{km}\right)$. To assess the robustness of our observed spectrum, we recalculate the zonal MSE spectrum over the same latitude range and time period using satellite data (CERES, light blue line). The CERES and ERA spectra agree very well at all wavelengths, although this agreement breaks down at short wavelengths ($\lambda<1,000\textnormal{km}$) if we do not average the data over one-day blocks (not shown). Hourly ERA data exhibit more spatial variability at short wavelengths, suggesting that CERES data and one-day time-averaging in the NG case may smooth out the MSE variability at short wavelengths.
Finally, \textcolor{black}{although the observational MSE spectrum flattens progressively at long wavelengths with no clear maximum}, spectra from idealized CPM exhibit a local maximum in the MSE variance at wavelengths of $\sim1,000-5,000\textnormal{km}$. \textcolor{black}{This peak is consistent with the strong $\sim5,000\textnormal{km}$ Madden-Julian Oscillation-like signal in NG CTRL and NG UNI-SEF \citep[][]{Khairoutdinov2018}, and the $\sim2,000\textnormal{km}$-long moist and dry regions of LC CTRL \citep[][]{Wing2016,Beucler2018d}. In observations, self-aggregation may not appear as a distinct peak in the MSE power spectrum, but simply as an enhancement of MSE variance over a broad range of scales. A peak might not form in the real Tropics for several reasons, including the larger amount of external forcing, lateral mixing, and amplifying diabatic feedbacks operating across a broader range of wavelengths. This motivates a quantitative framework to compare MSE tendencies across scales in models and observations: if the same processes enhance variance at large-scales, then self-aggregation processes likely play an important role in regulating observed MSE spectra.}

 \subsection{Adapting a Spectral Budget for Model-Observations Intercomparison}
 
 We now derive a formal spectral decomposition of the transient MSE budget terms to quantitatively assess the respective roles of separate processes in maintaining the spectrum at each wavelength. 

 This begins with the transient MSE budget, building on standard approaches, but modified so as to help fairly compare our simulations with observations, thus forging new ground. The transient MSE field $H^{\prime}\ $evolves in response to the net MSE flux at the atmospheric column's boundaries, with contributions from the net longwave flux $\dot{H}_{\mathrm{lw}}$, the net shortwave flux $\dot{H}_{\mathrm{sw}}$, the surface enthalpy fluxes $\dot{H}_{\mathrm{sf}}\ $and the advection of MSE through the column's boundaries $\dot{H}_{\mathrm{adv}}$. Separating the four MSE tendencies $\dot{H}_{i}\ $into their temporal mean $\overline{\dot{H}_{i}}\ $and their transient component $\dot{H}_{i}^{\prime}\ $ in the same spirit as equation \ref{eq:Transient_definition}, we can write the transient MSE budget as:

\begin{linenomath*}
\begin{equation}
\frac{\partial H^{\prime}}{\partial t}=\sum_{i=\mathrm{lw,sw,sf,adv}}\dot{H}_{i}^{\prime}.\label{eq:MSE_transient_budget}
\end{equation}
\end{linenomath*}
Following Section 2.2 of \citet{Beucler2018d}, we take the Fourier transform of Equation \ref{eq:MSE_transient_budget} and multiply it by the complex conjugate $\widehat{H^{\prime}}^{*}\ $of Equation \ref{eq:Fourier_transform} to derive a budget for the zonal spectrum $\varphi_{H}\ $of transient MSE:
\begin{linenomath*}
\begin{equation}
\frac{1}{2}\frac{\partial\varphi_{H}}{\partial t}=\sum_{i=\mathrm{lw,sw,sf,adv}}\Re\left(\widehat{H^{\prime}}^{*}\widehat{\dot{H}_{i}^{\prime}}\right),\label{eq:Spectral_MSE_budget}
\end{equation}
\end{linenomath*}
where $\Re\ $is the real part of a complex number. At each wavelength $\lambda$, MSE variance is created if a MSE tendency $\dot{H}_{i}\ $increases MSE where the transient MSE anomaly $H^{\prime}\ $is positive, corresponding to a positive co-spectrum $\Re\left(\widehat{H^{\prime}}^{*}\widehat{\dot{H}_{i}^{\prime}}\right)$. 
This framework generalizes the MSE variance framework of \citet{Bretherton2005} and \citet{Wing2014}, a particular case of Equation \ref{eq:Spectral_MSE_budget} that can be derived by integrating equation \ref{eq:Spectral_MSE_budget} across wavelengths before dividing it by the wavelength-integral of $\varphi_{H}\ $\citep[see Appendix B of ][]{Beucler2018d}. 
To yield an equation for the rate at which MSE tendencies maintain the MSE spectrum at each wavelength (in units $\textnormal{s}^{-1}$), we average equation \ref{eq:Spectral_MSE_budget} in time over a time-period $t_{\overline{H}}$ and divide it by the time-mean MSE spectrum $\overline{\varphi_{H}}\ $ at each wavelength:
\begin{linenomath*}
\begin{equation}
\frac{1}{t_{\overline{H}}}\frac{\Delta\varphi_{H}}{\overline{\varphi_{H}}}=\sum_{i=\mathrm{lw,sw,sf,adv}}\frac{2\overline{\Re\left(\widehat{H^{\prime}}^{*}\widehat{\dot{H}_{i}^{\prime}}\right)}}{\overline{\varphi_{H}}},\label{eq:Time-averaged-budget}
\end{equation}
\end{linenomath*}
where $\Delta\varphi_{H}\ $is the MSE spectrum difference between the beginning and the end of the time-average. We refer to the terms on the right-hand side of equation \ref{eq:Time-averaged-budget} as components of the spectral MSE variance tendency or, for brevity, as "variance rates".

Since Equation \ref{eq:Time-averaged-budget} \textcolor{black}{does not explicitly depend on} the time-mean zonal structure of the MSE tendencies, we can make direct analogies between observations and zonally-symmetric RCE, which is a key theoretical result of this paper. \textcolor{black}{Note that transient MSE tendencies themselves may be nonlinearly modulated by stationary features of low-level winds, MSE, clouds, among others. Therefore, equation \ref{eq:Time-averaged-budget} is not a closed theory for MSE transient variability; instead it provides a convenient diagnostic tool to compare the amount of variance injected by each MSE tendency $ \dot{H}_{i}$ across different base states}. The left-hand side of equation \ref{eq:Time-averaged-budget} is small \footnote{\textcolor{black}{The left-hand side of equation \ref{eq:Time-averaged-budget} equals 0.1\% of the longwave variance rate when the ERA dataset
is time-averaged over the Jan1,2010--Dec31,2014 period, 1.1\% of the
longwave variance rate when NG CTRL is time-averaged over 1 year,
and 5.1\% of the longwave variance rate when LC CTRL is time-averaged
over 40-80d.}} when the initial and final spectra $\varphi_{H}\ $are similar, or when the time-average is taken over a long time-period $t_{\overline{H}}$. 
\textcolor{black}{In both cases}, the four components of spectral MSE variance tendency on the right-hand side of equation \ref{eq:Time-averaged-budget} approximately balance. \textcolor{black}{Therefore, we can quantitatively compare the four rates of variance injection \textit{across scales} in models and observations.}


\subsection{Zonal Spectral Budget Intercomparison}

\begin{figure}[H]
\begin{centering}
\includegraphics[width=13cm]{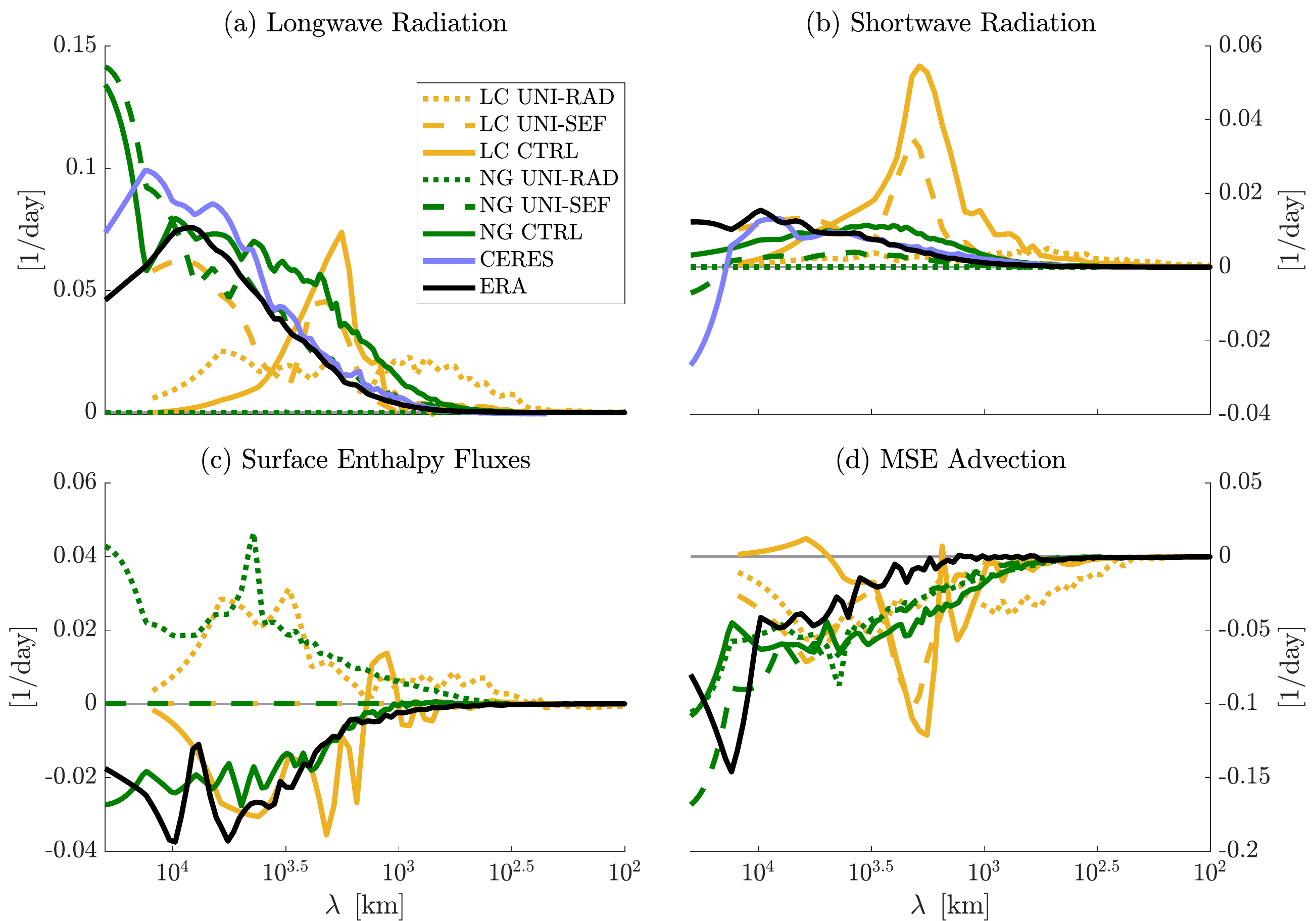}
\par\end{centering}

\caption{Rate at which (a) longwave radiation (b) shortwave radiation (c) surface enthalpy fluxes and (d) MSE advection maintain the MSE power spectrum at each wavelength (in units $\textnormal{day}^{-1}$) and for all datasets. The UNI-SEF rates of variance injection (dashed lines) have been divided by a factor of 5 because the denominator of equation \ref{eq:Time-averaged-budget}, which is the time-averaged spectrum $\overline{\varphi_{H}}\ $, is smaller for non-aggregated simulations. \textcolor{black}{Note the similar signs and shapes of the observed variance rates (light-blue and black lines) and the variance rates from \textit{aggregated} idealized simulations (yellow and green solid lines)}. \label{fig:MSE-spectral-rates}}
\end{figure}

The spectral rates of variance injection\textcolor{black}{,} depicted in Figure \ref{fig:MSE-spectral-rates}\textcolor{black}{,} have similar signs and amplitude across models \textcolor{black}{(green and yellow lines)} and observations \textcolor{black}{(black and light-blue lines)}. Surprisingly, even the LC \textcolor{black}{variance} rates (yellow lines) have similar signs and amplitudes to the \textcolor{black}{variance} rates from planetary-domain experiments, and are simply shifted to shorter wavelengths, despite the smaller zonal extent and 64:1 aspect ratio of the LC configuration. Therefore, we see the LC configuration as an idealized, reduced-size model to study the interaction between convection and the large-scale circulation, which makes LC a promising yet relatively inexpensive framework to study the processes maintaining convective aggregation across climates \citep{Wing2018}.

First, since longwave cooling to space is systematically lower in moist regions of high MSE \citep[e.g., ][]{Beucler2016b}, 
longwave radiation injects MSE variance at all wavelengths (Figure \ref{fig:MSE-spectral-rates}a), with rates as high at $1/\left(2\ \textnormal{weeks}\right)\ $at the planetary scale. Shortwave heating is larger in moister regions, mostly because of water vapor absorption \citep[e.g., sub-section 3.3 of ][]{Wing2017}, resulting in a shortwave injection of MSE variance at all wavelengths (Figure \ref{fig:MSE-spectral-rates}b). 
Surface enthalpy fluxes remove variance in observations and for idealized cases where convection has aggregated, while they unrealistically inject variance for the sensitivity tests (UNI-RAD) in which aggregation is artificially \textcolor{black}{prevented} (Figure \ref{fig:MSE-spectral-rates}c) or in the early phases of convective self-aggregation \citep[see Appendix D of ][]{Beucler2018d}. The difference between the rate at which surface fluxes remove variance in the aggregated and non-aggregated cases can be explained by decomposing the surface enthalpy fluxes into a wind-driven component and a component driven by the near-surface enthalpy disequilibrium. Section 3.4 of \citet{Beucler2018d} shows that while the wind-driven component favors convective aggregation (variance injection) because convective gustiness is higher is convectively-active regions, surface enthalpy disequilibrium is largest in dry regions, damping MSE variance (variance removal). As convection aggregates, MSE variance increases at long wavelengths and so does the surface enthalpy disequilibrium. In the real-world atmosphere, additional factors \textcolor{black}{such as higher near-surface wind speeds \citep[][]{Maloney2010}, ocean heat transport \citep[][]{Benedict2011}, and dry air intrusions \citep[][]{Bretherton2015a} can increase the smoothing effect of the disequilibrium-driven variability of surface enthalpy fluxes, while further decreasing the aggregating effect of the wind-driven variability}. This leads to larger surface flux damping at scales where radiation injects the most variance (Figure \ref{fig:MSE-spectral-rates}c), and \textcolor{black}{might contribute to} the absence of a peak in the ERA MSE spectrum (Figure \ref{fig:MSE_spectrum}b). MSE advection removes variance at all wavelengths with a maximum removal rate at the planetary scale (Figure \ref{fig:MSE-spectral-rates}d). \textcolor{black}{Since total advection is calculated as a residual of equation \ref{eq:Time-averaged-budget}, the fine variability of its variance rate may not be resolved, especially in ERA which does not close the MSE budget; explicitly calculating the horizontal and vertical components of MSE advection from three-dimensional data will be needed to clarify its scale-selectivity in observations and models.}

\section{Conclusion\label{sec:Conclusion}}

The multi-scale patterns of convective aggregation are directly connected to the hydrologic cycle in the Tropics  \citep[e.g., ][]{Kiranmayi2011}. While \textcolor{black}{convection-permitting} models have provided insight into the physical processes controlling convective aggregation, it has been hard to meaningfully \textcolor{black}{compare} idealized simulations against observations. We have addressed this issue by applying a spectral technique that reveals scale-selective aggregation processes in meteorological reanalyses, satellite retrievals, and idealized \textcolor{black}{convection-permitting} simulations of varying complexity.

The budget for the transient MSE spectrum exhibits scale-selective tendencies that hold across models and observations: longwave radiation injects variance at the longest wavelengths, shortwave radiation injects variance at long wavelengths, MSE advection removes variance across scales\textcolor{black}{, and} surface enthalpy fluxes mostly remove variance between $\lambda\approx1,000\textnormal{km}\ $and $\lambda\approx10,000\textnormal{km}$. We find a stronger damping effect of surface enthalpy fluxes in ERA reanalysis data relative to simulations that neglect ocean interaction and horizontal sea surface gradients. This finding is consistent with recent RCE simulations that have made surface flux feedbacks on aggregation more realistic by adding a meridional surface temperature gradient \citep[e.g., ][]{Bretherton2015a} or increasing surface temperature variability by adding a slab ocean \citep[e.g., ][]{Coppin2017,Hohenegger2016} or soil \citep[e.g., ][]{Hohenegger2018}, resulting in a damping of self-aggregation patterns. 

Removing the interaction between radiation and water vapor in the simulations prevents convective self-aggregation, resulting in a loss of MSE variance at long wavelengths ($\lambda>1,000\textnormal{km}$), and corresponding disagreement with the observed MSE variance. This adds to the growing body of evidence that radiatively-driven self-aggregation is key to generating realistic \textcolor{black}{moisture variability from homogeneous boundary conditions} \citep[e.g., ][]{Arnold2015}.


Undoubtedly, aspects of the causality are still murky since vertically-resolved, lower-tropospheric specific humidity, whose variance dominates the column MSE variance \textcolor{black}{\citep[][]{Holloway2009}}, may not directly respond to the thermodynamical constraints governing column MSE. For instance, is the longwave variance production peak too high for LC in Figure \ref{fig:MSE-spectral-rates}a because cloud-radiation processes are represented incorrectly, or because vertical advection of water vapor amplifies variance too much at a specific length scale?
\textcolor{black}{The} framework introduced here generalizes to three-dimensional tracer variance budgets, and could be used to investigate the processes injecting zonal variance in the lower-tropospheric water vapor spectrum at long wavelengths.

Ultimately, we hope the tool summarized here can be deployed across the emerging hierarchy of global cloud resolving models \citep{Satoh2019} to help clarify their intrinsic thermodynamics. \textcolor{black}{Our spectral framework can be generalized to spatially-limited domains by choosing a transform insensitive to non-periodic boundaries, such as the discrete cosine transform \citep[e.g., ][]{Denis2002,Selz2018}. It can also be generalized to arbitrary subsets of the domain by choosing a transform retaining localization information, such as the wavelet transform \citep[e.g., ][]{Torrence1998}}. While spatio-temporal spectra are familiar to tropical dynamicists \citep[e.g., ][]{Wheeler1999,Yasunaga2019}, formal spectral decomposition of underlying process budgets are not yet in widespread use. In this context, traditional diagnostic tools may fail to compactly analyze the underlying causes of multi-scale discrepancies across models. By quantifying the preferential scales of zonal thermodynamic variability, our spectral framework allows comparison between models and observational datasets across configurations, resolutions, and scales. 

\acknowledgments
Tom Beucler is supported by NSF grants AGS-1520683 and OAC-1835769, Tristan Abbott and Timothy Cronin are supported by NSF grants AGS-1740533 and AGS-1623218, and Mike Pritchard is supported by NSF grant AGS-1734164 and DOE grant DE-SC0012152. We thank Kerry Emanuel, Paul O'Gorman, Zhiming Kuang and Chris Bretherton
for review and guidance on an early version of this manuscript, \textcolor{black}{and two anonymous reviewers who helped improve the quality of the present manuscript}. The source code and data used to produce the figures can be found
at \url{https://github.com/tbeucler/2019\_Convective\_SA\_MSE\_Transients}. The ERA reanalysis data was downloaded from the Copernicus Data Store,
the CERES data was downloaded from the CERES NASA website, the NG
data is stored on the Cheyenne computing cluster provided by NCAR,
and the LC data is stored on the Engaging computing cluster provided
by MIT.


%
\bibliography{My_Collection.bib}

\begin{thebibliography}{}

\bibitem [\protect \citeauthoryear {%
Arnold%
\ \BBA {} Randall%
}{%
Arnold%
\ \BBA {} Randall%
}{%
{\protect \APACyear {2015}}%
}]{%
Arnold2015}
\APACinsertmetastar {%
Arnold2015}%
\begin{APACrefauthors}%
Arnold, N\BPBI P.%
\BCBT {}\ \BBA {} Randall, D\BPBI A.%
\end{APACrefauthors}%
\unskip\
\newblock
\APACrefYearMonthDay{2015}{dec}{}.
\newblock
{\BBOQ}\APACrefatitle {{Global-scale convective aggregation: Implications for
  the Madden-Julian Oscillation}} {{Global-scale convective aggregation:
  Implications for the Madden-Julian Oscillation}}.{\BBCQ}
\newblock
\APACjournalVolNumPages{Journal of Advances in Modeling Earth
  Systems}{7}{4}{1499--1518}.
\newblock
\begin{APACrefURL} \url{http://doi.wiley.com/10.1002/2015MS000498}
  \end{APACrefURL}
\newblock
\begin{APACrefDOI} \doi{10.1002/2015MS000498} \end{APACrefDOI}
\PrintBackRefs{\CurrentBib}

\bibitem [\protect \citeauthoryear {%
Benedict%
\ \BBA {} Randall%
}{%
Benedict%
\ \BBA {} Randall%
}{%
{\protect \APACyear {2011}}%
}]{%
Benedict2011}
\APACinsertmetastar {%
Benedict2011}%
\begin{APACrefauthors}%
Benedict, J\BPBI J.%
\BCBT {}\ \BBA {} Randall, D\BPBI A.%
\end{APACrefauthors}%
\unskip\
\newblock
\APACrefYearMonthDay{2011}{sep}{}.
\newblock
{\BBOQ}\APACrefatitle {{Impacts of Idealized Air-Sea Coupling on Madden-Julian
  Oscillation Structure in the Superparameterized CAM}} {{Impacts of Idealized
  Air-Sea Coupling on Madden-Julian Oscillation Structure in the
  Superparameterized CAM}}.{\BBCQ}
\newblock
\APACjournalVolNumPages{Journal of the Atmospheric
  Sciences}{68}{9}{1990--2008}.
\newblock
\begin{APACrefURL}
  \url{http://journals.ametsoc.org/doi/abs/10.1175/JAS-D-11-04.1}
  \end{APACrefURL}
\newblock
\begin{APACrefDOI} \doi{10.1175/jas-d-11-04.1} \end{APACrefDOI}
\PrintBackRefs{\CurrentBib}

\bibitem [\protect \citeauthoryear {%
Beucher%
, Lafore%
, Karbou%
\BCBL {}\ \BBA {} Roca%
}{%
Beucher%
\ \protect \BOthers {.}}{%
{\protect \APACyear {2014}}%
}]{%
Beucher2014}
\APACinsertmetastar {%
Beucher2014}%
\begin{APACrefauthors}%
Beucher, F.%
, Lafore, J\BPBI P.%
, Karbou, F.%
\BCBL {}\ \BBA {} Roca, R.%
\end{APACrefauthors}%
\unskip\
\newblock
\APACrefYearMonthDay{2014}{jul}{}.
\newblock
{\BBOQ}\APACrefatitle {{High-resolution prediction of a major convective period
  over West Africa}} {{High-resolution prediction of a major convective period
  over West Africa}}.{\BBCQ}
\newblock
\APACjournalVolNumPages{Quarterly Journal of the Royal Meteorological
  Society}{140}{682}{1409--1425}.
\newblock
\begin{APACrefURL} \url{http://doi.wiley.com/10.1002/qj.2225} \end{APACrefURL}
\newblock
\begin{APACrefDOI} \doi{10.1002/qj.2225} \end{APACrefDOI}
\PrintBackRefs{\CurrentBib}

\bibitem [\protect \citeauthoryear {%
Beucler%
\ \BBA {} Cronin%
}{%
Beucler%
\ \BBA {} Cronin%
}{%
{\protect \APACyear {2018}}%
}]{%
Beucler2018d}
\APACinsertmetastar {%
Beucler2018d}%
\begin{APACrefauthors}%
Beucler, T.%
\BCBT {}\ \BBA {} Cronin, T.%
\end{APACrefauthors}%
\unskip\
\newblock
\APACrefYearMonthDay{2018}{feb}{}.
\newblock
{\BBOQ}\APACrefatitle {{A Budget for the Size of Convective Self-aggregation}}
  {{A Budget for the Size of Convective Self-aggregation}}.{\BBCQ}
\newblock
\APACjournalVolNumPages{Quarterly Journal of the Royal Meteorological
  Society}{}{}{}.
\newblock
\begin{APACrefURL} \url{http://doi.wiley.com/10.1002/qj.3468} \end{APACrefURL}
\newblock
\begin{APACrefDOI} \doi{10.1002/qj.3468} \end{APACrefDOI}
\PrintBackRefs{\CurrentBib}

\bibitem [\protect \citeauthoryear {%
Beucler%
\ \BBA {} Cronin%
}{%
Beucler%
\ \BBA {} Cronin%
}{%
{\protect \APACyear {2016}}%
}]{%
Beucler2016b}
\APACinsertmetastar {%
Beucler2016b}%
\begin{APACrefauthors}%
Beucler, T.%
\BCBT {}\ \BBA {} Cronin, T\BPBI W.%
\end{APACrefauthors}%
\unskip\
\newblock
\APACrefYearMonthDay{2016}{dec}{}.
\newblock
\APACrefbtitle {{Moisture-radiative cooling instability}} {{Moisture-radiative
  cooling instability}}\ (\BVOL~8)\ (\BNUM~4).
\newblock
\begin{APACrefURL} \url{http://doi.wiley.com/10.1002/2016MS000763}
  \end{APACrefURL}
\newblock
\begin{APACrefDOI} \doi{10.1002/2016MS000763} \end{APACrefDOI}
\PrintBackRefs{\CurrentBib}

\bibitem [\protect \citeauthoryear {%
Bretherton%
, Blossey%
\BCBL {}\ \BBA {} Khairoutdinov%
}{%
Bretherton%
\ \protect \BOthers {.}}{%
{\protect \APACyear {2005}}%
}]{%
Bretherton2005}
\APACinsertmetastar {%
Bretherton2005}%
\begin{APACrefauthors}%
Bretherton, C\BPBI S.%
, Blossey, P\BPBI N.%
\BCBL {}\ \BBA {} Khairoutdinov, M.%
\end{APACrefauthors}%
\unskip\
\newblock
\APACrefYearMonthDay{2005}{}{}.
\newblock
{\BBOQ}\APACrefatitle {{An Energy-Balance Analysis of Deep Convective
  Self-Aggregation above Uniform SST}} {{An Energy-Balance Analysis of Deep
  Convective Self-Aggregation above Uniform SST}}.{\BBCQ}
\newblock
\APACjournalVolNumPages{Journal of the Atmospheric
  Sciences}{62}{12}{4273--4292}.
\newblock
\begin{APACrefDOI} \doi{10.1175/JAS3614.1} \end{APACrefDOI}
\PrintBackRefs{\CurrentBib}

\bibitem [\protect \citeauthoryear {%
Bretherton%
\ \BBA {} Khairoutdinov%
}{%
Bretherton%
\ \BBA {} Khairoutdinov%
}{%
{\protect \APACyear {2015}}%
}]{%
Bretherton2015a}
\APACinsertmetastar {%
Bretherton2015a}%
\begin{APACrefauthors}%
Bretherton, C\BPBI S.%
\BCBT {}\ \BBA {} Khairoutdinov, M\BPBI F.%
\end{APACrefauthors}%
\unskip\
\newblock
\APACrefYearMonthDay{2015}{dec}{}.
\newblock
{\BBOQ}\APACrefatitle {{Convective self-aggregation feedbacks in near-global
  cloud-resolving simulations of an aquaplanet}} {{Convective self-aggregation
  feedbacks in near-global cloud-resolving simulations of an
  aquaplanet}}.{\BBCQ}
\newblock
\APACjournalVolNumPages{Journal of Advances in Modeling Earth
  Systems}{7}{4}{1765--1787}.
\newblock
\begin{APACrefURL} \url{http://doi.wiley.com/10.1002/2015MS000499}
  \end{APACrefURL}
\newblock
\begin{APACrefDOI} \doi{10.1002/2015MS000499} \end{APACrefDOI}
\PrintBackRefs{\CurrentBib}

\bibitem [\protect \citeauthoryear {%
Coppin%
\ \BBA {} Bony%
}{%
Coppin%
\ \BBA {} Bony%
}{%
{\protect \APACyear {2017}}%
}]{%
Coppin2017}
\APACinsertmetastar {%
Coppin2017}%
\begin{APACrefauthors}%
Coppin, D.%
\BCBT {}\ \BBA {} Bony, S.%
\end{APACrefauthors}%
\unskip\
\newblock
\APACrefYearMonthDay{2017}{may}{}.
\newblock
{\BBOQ}\APACrefatitle {{Internal variability in a coupled General Circulation
  Model in Radiative-Convective Equilibrium}} {{Internal variability in a
  coupled General Circulation Model in Radiative-Convective
  Equilibrium}}.{\BBCQ}
\newblock
\APACjournalVolNumPages{Geophysical Research Letters}{}{}{}.
\newblock
\begin{APACrefURL} \url{http://doi.wiley.com/10.1002/2017GL073658}
  \end{APACrefURL}
\newblock
\begin{APACrefDOI} \doi{10.1002/2017GL073658} \end{APACrefDOI}
\PrintBackRefs{\CurrentBib}

\bibitem [\protect \citeauthoryear {%
Denis%
, C{\^{o}}t{\'{e}}%
\BCBL {}\ \BBA {} Laprise%
}{%
Denis%
\ \protect \BOthers {.}}{%
{\protect \APACyear {2002}}%
}]{%
Denis2002}
\APACinsertmetastar {%
Denis2002}%
\begin{APACrefauthors}%
Denis, B.%
, C{\^{o}}t{\'{e}}, J.%
\BCBL {}\ \BBA {} Laprise, R.%
\end{APACrefauthors}%
\unskip\
\newblock
\APACrefYearMonthDay{2002}{jul}{}.
\newblock
{\BBOQ}\APACrefatitle {{Spectral Decomposition of Two-Dimensional Atmospheric
  Fields on Limited-Area Domains Using the Discrete Cosine Transform (DCT)}}
  {{Spectral Decomposition of Two-Dimensional Atmospheric Fields on
  Limited-Area Domains Using the Discrete Cosine Transform (DCT)}}.{\BBCQ}
\newblock
\APACjournalVolNumPages{Monthly Weather Review}{130}{7}{1812--1829}.
\newblock
\begin{APACrefURL}
  \url{http://journals.ametsoc.org/doi/abs/10.1175/1520-0493{\%}282002{\%}29130{\%}3C1812{\%}3ASDOTDA{\%}3E2.0.CO{\%}3B2}
  \end{APACrefURL}
\newblock
\begin{APACrefDOI} \doi{10.1175/1520-0493(2002)130<1812:sdotda>2.0.co;2}
  \end{APACrefDOI}
\PrintBackRefs{\CurrentBib}

\bibitem [\protect \citeauthoryear {%
Feldl%
, Frierson%
\BCBL {}\ \BBA {} Roe%
}{%
Feldl%
\ \protect \BOthers {.}}{%
{\protect \APACyear {2014}}%
}]{%
Feldl2014}
\APACinsertmetastar {%
Feldl2014}%
\begin{APACrefauthors}%
Feldl, N.%
, Frierson, D\BPBI M\BPBI W.%
\BCBL {}\ \BBA {} Roe, G\BPBI H.%
\end{APACrefauthors}%
\unskip\
\newblock
\APACrefYearMonthDay{2014}{mar}{}.
\newblock
{\BBOQ}\APACrefatitle {{The influence of regional feedbacks on circulation
  sensitivity}} {{The influence of regional feedbacks on circulation
  sensitivity}}.{\BBCQ}
\newblock
\APACjournalVolNumPages{Geophysical Research Letters}{41}{6}{2212--2220}.
\newblock
\begin{APACrefURL} \url{http://doi.wiley.com/10.1002/2014GL059336}
  \end{APACrefURL}
\newblock
\begin{APACrefDOI} \doi{10.1002/2014GL059336} \end{APACrefDOI}
\PrintBackRefs{\CurrentBib}

\bibitem [\protect \citeauthoryear {%
Frigo%
\ \BBA {} Johnson%
}{%
Frigo%
\ \BBA {} Johnson%
}{%
{\protect \APACyear {2005}}%
}]{%
Frigo2005}
\APACinsertmetastar {%
Frigo2005}%
\begin{APACrefauthors}%
Frigo, M.%
\BCBT {}\ \BBA {} Johnson, S\BPBI G.%
\end{APACrefauthors}%
\unskip\
\newblock
\APACrefYearMonthDay{2005}{}{}.
\newblock
{\BBOQ}\APACrefatitle {{The design and implementation of FFTW3}} {{The design
  and implementation of FFTW3}}.{\BBCQ}
\newblock
\BIn{} \APACrefbtitle {Proceedings of the IEEE} {Proceedings of the ieee}\
  (\BVOL~93, \BPGS\ 216--231).
\newblock
\begin{APACrefURL} \url{http://www.fftw.org/fftw-paper-ieee.pdf}
  \end{APACrefURL}
\newblock
\begin{APACrefDOI} \doi{10.1109/JPROC.2004.840301} \end{APACrefDOI}
\PrintBackRefs{\CurrentBib}

\bibitem [\protect \citeauthoryear {%
Held%
, Hemler%
\BCBL {}\ \BBA {} Ramaswamy%
}{%
Held%
\ \protect \BOthers {.}}{%
{\protect \APACyear {1993}}%
}]{%
Held1993}
\APACinsertmetastar {%
Held1993}%
\begin{APACrefauthors}%
Held, I\BPBI M.%
, Hemler, R\BPBI S.%
\BCBL {}\ \BBA {} Ramaswamy, V.%
\end{APACrefauthors}%
\unskip\
\newblock
\APACrefYearMonthDay{1993}{}{}.
\newblock
\APACrefbtitle {{Radiative-Convective Equilibrium with Explicit Two-Dimensional
  Moist Convection}} {{Radiative-Convective Equilibrium with Explicit
  Two-Dimensional Moist Convection}}\ (\BVOL~50)\ (\BNUM~23).
\newblock
\begin{APACrefDOI} \doi{10.1175/1520-0469(1993)050<3909:RCEWET>2.0.CO;2}
  \end{APACrefDOI}
\PrintBackRefs{\CurrentBib}

\bibitem [\protect \citeauthoryear {%
Hersbach%
\ \BBA {} H.%
}{%
Hersbach%
\ \BBA {} H.%
}{%
{\protect \APACyear {2016}}%
}]{%
Hersbach2016}
\APACinsertmetastar {%
Hersbach2016}%
\begin{APACrefauthors}%
Hersbach, H.%
\BCBT {}\ \BBA {} H.%
\end{APACrefauthors}%
\unskip\
\newblock
\APACrefYearMonthDay{2016}{}{}.
\newblock
{\BBOQ}\APACrefatitle {{The ERA5 Atmospheric Reanalysis.}} {{The ERA5
  Atmospheric Reanalysis.}}{\BBCQ}
\newblock
\APACjournalVolNumPages{American Geophysical Union, Fall General Assembly 2016,
  abstract id. NG33D-01}{}{}{}.
\newblock
\begin{APACrefURL} \url{http://adsabs.harvard.edu/abs/2016AGUFMNG33D..01H}
  \end{APACrefURL}
\PrintBackRefs{\CurrentBib}

\bibitem [\protect \citeauthoryear {%
Hill%
\ \BBA {} Lackmann%
}{%
Hill%
\ \BBA {} Lackmann%
}{%
{\protect \APACyear {2009}}%
}]{%
Hill2009}
\APACinsertmetastar {%
Hill2009}%
\begin{APACrefauthors}%
Hill, K\BPBI A.%
\BCBT {}\ \BBA {} Lackmann, G\BPBI M.%
\end{APACrefauthors}%
\unskip\
\newblock
\APACrefYearMonthDay{2009}{oct}{}.
\newblock
{\BBOQ}\APACrefatitle {{Influence of Environmental Humidity on Tropical Cyclone
  Size}} {{Influence of Environmental Humidity on Tropical Cyclone
  Size}}.{\BBCQ}
\newblock
\APACjournalVolNumPages{Monthly Weather Review}{137}{10}{3294--3315}.
\newblock
\begin{APACrefURL}
  \url{http://journals.ametsoc.org/doi/abs/10.1175/2009MWR2679.1}
  \end{APACrefURL}
\newblock
\begin{APACrefDOI} \doi{10.1175/2009MWR2679.1} \end{APACrefDOI}
\PrintBackRefs{\CurrentBib}

\bibitem [\protect \citeauthoryear {%
Hohenegger%
\ \BBA {} Stevens%
}{%
Hohenegger%
\ \BBA {} Stevens%
}{%
{\protect \APACyear {2016}}%
}]{%
Hohenegger2016}
\APACinsertmetastar {%
Hohenegger2016}%
\begin{APACrefauthors}%
Hohenegger, C.%
\BCBT {}\ \BBA {} Stevens, B.%
\end{APACrefauthors}%
\unskip\
\newblock
\APACrefYearMonthDay{2016}{sep}{}.
\newblock
{\BBOQ}\APACrefatitle {{Coupled radiative convective equilibrium simulations
  with explicit and parameterized convection}} {{Coupled radiative convective
  equilibrium simulations with explicit and parameterized convection}}.{\BBCQ}
\newblock
\APACjournalVolNumPages{Journal of Advances in Modeling Earth
  Systems}{8}{3}{1468--1482}.
\newblock
\begin{APACrefURL} \url{http://doi.wiley.com/10.1002/2016MS000666}
  \end{APACrefURL}
\newblock
\begin{APACrefDOI} \doi{10.1002/2016MS000666} \end{APACrefDOI}
\PrintBackRefs{\CurrentBib}

\bibitem [\protect \citeauthoryear {%
Hohenegger%
\ \BBA {} Stevens%
}{%
Hohenegger%
\ \BBA {} Stevens%
}{%
{\protect \APACyear {2018}}%
}]{%
Hohenegger2018}
\APACinsertmetastar {%
Hohenegger2018}%
\begin{APACrefauthors}%
Hohenegger, C.%
\BCBT {}\ \BBA {} Stevens, B.%
\end{APACrefauthors}%
\unskip\
\newblock
\APACrefYearMonthDay{2018}{may}{}.
\newblock
{\BBOQ}\APACrefatitle {{The role of the permanent wilting point in controlling
  the spatial distribution of precipitation}} {{The role of the permanent
  wilting point in controlling the spatial distribution of
  precipitation}}.{\BBCQ}
\newblock
\APACjournalVolNumPages{Proceedings of the National Academy of
  Sciences}{115}{22}{5692--5697}.
\newblock
\begin{APACrefURL} \url{http://www.ncbi.nlm.nih.gov/pubmed/29760083
  http://www.pubmedcentral.nih.gov/articlerender.fcgi?artid=PMC5984498
  http://www.pnas.org/lookup/doi/10.1073/pnas.1718842115} \end{APACrefURL}
\newblock
\begin{APACrefDOI} \doi{10.1073/pnas.1718842115} \end{APACrefDOI}
\PrintBackRefs{\CurrentBib}

\bibitem [\protect \citeauthoryear {%
Holloway%
}{%
Holloway%
}{%
{\protect \APACyear {2017}}%
}]{%
Holloway2017a}
\APACinsertmetastar {%
Holloway2017a}%
\begin{APACrefauthors}%
Holloway, C\BPBI E.%
\end{APACrefauthors}%
\unskip\
\newblock
\APACrefYearMonthDay{2017}{jun}{}.
\newblock
{\BBOQ}\APACrefatitle {{Convective aggregation in realistic convective-scale
  simulations}} {{Convective aggregation in realistic convective-scale
  simulations}}.{\BBCQ}
\newblock
\APACjournalVolNumPages{Journal of Advances in Modeling Earth
  Systems}{9}{2}{1450--1472}.
\newblock
\begin{APACrefURL} \url{http://doi.wiley.com/10.1002/2017MS000980}
  \end{APACrefURL}
\newblock
\begin{APACrefDOI} \doi{10.1002/2017MS000980} \end{APACrefDOI}
\PrintBackRefs{\CurrentBib}

\bibitem [\protect \citeauthoryear {%
Holloway%
\ \BBA {} Neelin%
}{%
Holloway%
\ \BBA {} Neelin%
}{%
{\protect \APACyear {2009}}%
}]{%
Holloway2009}
\APACinsertmetastar {%
Holloway2009}%
\begin{APACrefauthors}%
Holloway, C\BPBI E.%
\BCBT {}\ \BBA {} Neelin, J\BPBI D.%
\end{APACrefauthors}%
\unskip\
\newblock
\APACrefYearMonthDay{2009}{jun}{}.
\newblock
{\BBOQ}\APACrefatitle {{Moisture Vertical Structure, Column Water Vapor, and
  Tropical Deep Convection}} {{Moisture Vertical Structure, Column Water Vapor,
  and Tropical Deep Convection}}.{\BBCQ}
\newblock
\APACjournalVolNumPages{Journal of the Atmospheric
  Sciences}{66}{6}{1665--1683}.
\newblock
\begin{APACrefURL}
  \url{http://journals.ametsoc.org/doi/abs/10.1175/2008JAS2806.1
  http://dx.doi.org/10.1175/2008JAS2806.1} \end{APACrefURL}
\newblock
\begin{APACrefDOI} \doi{10.1175/2008JAS2806.1} \end{APACrefDOI}
\PrintBackRefs{\CurrentBib}

\bibitem [\protect \citeauthoryear {%
Holloway%
\ \protect \BOthers {.}}{%
Holloway%
\ \protect \BOthers {.}}{%
{\protect \APACyear {2017}}%
}]{%
Holloway2017}
\APACinsertmetastar {%
Holloway2017}%
\begin{APACrefauthors}%
Holloway, C\BPBI E.%
, Wing, A.%
, Bony, S.%
, Muller, C.%
, Masunaga, H.%
, L'Ecuyer, T\BPBI S.%
\BDBL {}Zuidema, P.%
\end{APACrefauthors}%
\unskip\
\newblock
\APACrefYearMonthDay{2017}{nov}{}.
\newblock
\APACrefbtitle {{Observing Convective Aggregation}} {{Observing Convective
  Aggregation}}\ (\BVOL~38)\ (\BNUM~6).
\newblock
\APACaddressPublisher{}{Springer Netherlands}.
\newblock
\begin{APACrefURL} \url{http://link.springer.com/10.1007/s10712-017-9419-1}
  \end{APACrefURL}
\newblock
\begin{APACrefDOI} \doi{10.1007/s10712-017-9419-1} \end{APACrefDOI}
\PrintBackRefs{\CurrentBib}

\bibitem [\protect \citeauthoryear {%
Houze%
}{%
Houze%
}{%
{\protect \APACyear {2004}}%
}]{%
Houze2004}
\APACinsertmetastar {%
Houze2004}%
\begin{APACrefauthors}%
Houze, R\BPBI A.%
\end{APACrefauthors}%
\unskip\
\newblock
\APACrefYearMonthDay{2004}{}{}.
\newblock
\APACrefbtitle {{Mesoscale convective systems}} {{Mesoscale convective
  systems}}\ (\BVOL~42)\ (\BNUM~4).
\newblock
\begin{APACrefURL} \url{http://doi.wiley.com/10.1029/2004RG000150}
  \end{APACrefURL}
\newblock
\begin{APACrefDOI} \doi{10.1029/2004RG000150} \end{APACrefDOI}
\PrintBackRefs{\CurrentBib}

\bibitem [\protect \citeauthoryear {%
Khairoutdinov%
\ \BBA {} Emanuel%
}{%
Khairoutdinov%
\ \BBA {} Emanuel%
}{%
{\protect \APACyear {2018}}%
}]{%
Khairoutdinov2018}
\APACinsertmetastar {%
Khairoutdinov2018}%
\begin{APACrefauthors}%
Khairoutdinov, M\BPBI F.%
\BCBT {}\ \BBA {} Emanuel, K.%
\end{APACrefauthors}%
\unskip\
\newblock
\APACrefYearMonthDay{2018}{oct}{}.
\newblock
{\BBOQ}\APACrefatitle {{Intraseasonal Variability in a Cloud-Permitting
  Near-Global Equatorial Aqua-Planet Model}} {{Intraseasonal Variability in a
  Cloud-Permitting Near-Global Equatorial Aqua-Planet Model}}.{\BBCQ}
\newblock
\APACjournalVolNumPages{Journal of the Atmospheric
  Sciences}{}{}{JAS--D--18--0152.1}.
\newblock
\begin{APACrefURL}
  \url{http://journals.ametsoc.org/doi/10.1175/JAS-D-18-0152.1}
  \end{APACrefURL}
\newblock
\begin{APACrefDOI} \doi{10.1175/JAS-D-18-0152.1} \end{APACrefDOI}
\PrintBackRefs{\CurrentBib}

\bibitem [\protect \citeauthoryear {%
Khairoutdinov%
\ \BBA {} Randall%
}{%
Khairoutdinov%
\ \BBA {} Randall%
}{%
{\protect \APACyear {2003}}%
}]{%
Khairoutdinov2003}
\APACinsertmetastar {%
Khairoutdinov2003}%
\begin{APACrefauthors}%
Khairoutdinov, M\BPBI F.%
\BCBT {}\ \BBA {} Randall, D\BPBI a.%
\end{APACrefauthors}%
\unskip\
\newblock
\APACrefYearMonthDay{2003}{}{}.
\newblock
{\BBOQ}\APACrefatitle {{Cloud Resolving Modeling of the ARM Summer 1997 IOP:
  Model Formulation, Results, Uncertainties, and Sensitivities}} {{Cloud
  Resolving Modeling of the ARM Summer 1997 IOP: Model Formulation, Results,
  Uncertainties, and Sensitivities}}.{\BBCQ}
\newblock
\APACjournalVolNumPages{Journal of the Atmospheric Sciences}{60}{}{607--625}.
\newblock
\begin{APACrefDOI} \doi{10.1175/1520-0469(2003)060<0607:CRMOTA>2.0.CO;2}
  \end{APACrefDOI}
\PrintBackRefs{\CurrentBib}

\bibitem [\protect \citeauthoryear {%
Kiranmayi%
\ \BBA {} Maloney%
}{%
Kiranmayi%
\ \BBA {} Maloney%
}{%
{\protect \APACyear {2011}}%
}]{%
Kiranmayi2011}
\APACinsertmetastar {%
Kiranmayi2011}%
\begin{APACrefauthors}%
Kiranmayi, L.%
\BCBT {}\ \BBA {} Maloney, E\BPBI D.%
\end{APACrefauthors}%
\unskip\
\newblock
\APACrefYearMonthDay{2011}{nov}{}.
\newblock
{\BBOQ}\APACrefatitle {{Intraseasonal moist static energy budget in reanalysis
  data}} {{Intraseasonal moist static energy budget in reanalysis
  data}}.{\BBCQ}
\newblock
\APACjournalVolNumPages{Journal of Geophysical Research
  Atmospheres}{116}{21}{}.
\newblock
\begin{APACrefURL} \url{http://doi.wiley.com/10.1029/2011JD016031}
  \end{APACrefURL}
\newblock
\begin{APACrefDOI} \doi{10.1029/2011JD016031} \end{APACrefDOI}
\PrintBackRefs{\CurrentBib}

\bibitem [\protect \citeauthoryear {%
Laing%
, Trier%
\BCBL {}\ \BBA {} Davis%
}{%
Laing%
\ \protect \BOthers {.}}{%
{\protect \APACyear {2012}}%
}]{%
Laing2012}
\APACinsertmetastar {%
Laing2012}%
\begin{APACrefauthors}%
Laing, A\BPBI G.%
, Trier, S\BPBI B.%
\BCBL {}\ \BBA {} Davis, C\BPBI A.%
\end{APACrefauthors}%
\unskip\
\newblock
\APACrefYearMonthDay{2012}{sep}{}.
\newblock
{\BBOQ}\APACrefatitle {{Numerical Simulation of Episodes of Organized
  Convection in Tropical Northern Africa}} {{Numerical Simulation of Episodes
  of Organized Convection in Tropical Northern Africa}}.{\BBCQ}
\newblock
\APACjournalVolNumPages{Monthly Weather Review}{140}{9}{2874--2886}.
\newblock
\begin{APACrefURL}
  \url{http://journals.ametsoc.org/doi/abs/10.1175/MWR-D-11-00330.1}
  \end{APACrefURL}
\newblock
\begin{APACrefDOI} \doi{10.1175/mwr-d-11-00330.1} \end{APACrefDOI}
\PrintBackRefs{\CurrentBib}

\bibitem [\protect \citeauthoryear {%
LeMone%
, Zipser%
\BCBL {}\ \BBA {} Trier%
}{%
LeMone%
\ \protect \BOthers {.}}{%
{\protect \APACyear {1998}}%
}]{%
LeMone1998}
\APACinsertmetastar {%
LeMone1998}%
\begin{APACrefauthors}%
LeMone, M\BPBI A.%
, Zipser, E\BPBI J.%
\BCBL {}\ \BBA {} Trier, S\BPBI B.%
\end{APACrefauthors}%
\unskip\
\newblock
\APACrefYearMonthDay{1998}{dec}{}.
\newblock
{\BBOQ}\APACrefatitle {{The Role of Environmental Shear and Thermodynamic
  Conditions in Determining the Structure and Evolution of Mesoscale Convective
  Systems during TOGA COARE}} {{The Role of Environmental Shear and
  Thermodynamic Conditions in Determining the Structure and Evolution of
  Mesoscale Convective Systems during TOGA COARE}}.{\BBCQ}
\newblock
\APACjournalVolNumPages{Journal of the Atmospheric
  Sciences}{55}{23}{3493--3518}.
\newblock
\begin{APACrefURL}
  \url{http://journals.ametsoc.org/doi/abs/10.1175/1520-0469{\%}281998{\%}29055{\%}3C3493{\%}3ATROESA{\%}3E2.0.CO{\%}3B2}
  \end{APACrefURL}
\newblock
\begin{APACrefDOI} \doi{10.1175/1520-0469(1998)055<3493:TROESA>2.0.CO;2}
  \end{APACrefDOI}
\PrintBackRefs{\CurrentBib}

\bibitem [\protect \citeauthoryear {%
Maloney%
, Sobel%
\BCBL {}\ \BBA {} Hannah%
}{%
Maloney%
\ \protect \BOthers {.}}{%
{\protect \APACyear {2010}}%
}]{%
Maloney2010}
\APACinsertmetastar {%
Maloney2010}%
\begin{APACrefauthors}%
Maloney, E\BPBI D.%
, Sobel, A\BPBI H.%
\BCBL {}\ \BBA {} Hannah, W\BPBI M.%
\end{APACrefauthors}%
\unskip\
\newblock
\APACrefYearMonthDay{2010}{apr}{}.
\newblock
{\BBOQ}\APACrefatitle {{Intraseasonal variability in an aquaplanet general
  circulation model}} {{Intraseasonal variability in an aquaplanet general
  circulation model}}.{\BBCQ}
\newblock
\APACjournalVolNumPages{Journal of Advances in Modeling Earth
  Systems}{2}{2}{5}.
\newblock
\begin{APACrefURL} \url{http://doi.wiley.com/10.3894/JAMES.2010.2.5}
  \end{APACrefURL}
\newblock
\begin{APACrefDOI} \doi{10.3894/JAMES.2010.2.5} \end{APACrefDOI}
\PrintBackRefs{\CurrentBib}

\bibitem [\protect \citeauthoryear {%
Montgomery%
\ \BBA {} Smith%
}{%
Montgomery%
\ \BBA {} Smith%
}{%
{\protect \APACyear {2017}}%
}]{%
Montgomery2017}
\APACinsertmetastar {%
Montgomery2017}%
\begin{APACrefauthors}%
Montgomery, M\BPBI T.%
\BCBT {}\ \BBA {} Smith, R\BPBI K.%
\end{APACrefauthors}%
\unskip\
\newblock
\APACrefYearMonthDay{2017}{jan}{}.
\newblock
{\BBOQ}\APACrefatitle {{Recent Developments in the Fluid Dynamics of Tropical
  Cyclones}} {{Recent Developments in the Fluid Dynamics of Tropical
  Cyclones}}.{\BBCQ}
\newblock
\APACjournalVolNumPages{Annual Review of Fluid Mechanics}{49}{1}{541--574}.
\newblock
\begin{APACrefURL}
  \url{http://www.annualreviews.org/doi/10.1146/annurev-fluid-010816-060022}
  \end{APACrefURL}
\newblock
\begin{APACrefDOI} \doi{10.1146/annurev-fluid-010816-060022} \end{APACrefDOI}
\PrintBackRefs{\CurrentBib}

\bibitem [\protect \citeauthoryear {%
Muller%
\ \BBA {} Bony%
}{%
Muller%
\ \BBA {} Bony%
}{%
{\protect \APACyear {2015}}%
}]{%
Muller2015a}
\APACinsertmetastar {%
Muller2015a}%
\begin{APACrefauthors}%
Muller, C.%
\BCBT {}\ \BBA {} Bony, S.%
\end{APACrefauthors}%
\unskip\
\newblock
\APACrefYearMonthDay{2015}{jul}{}.
\newblock
{\BBOQ}\APACrefatitle {{What favors convective aggregation and why?}} {{What
  favors convective aggregation and why?}}{\BBCQ}
\newblock
\APACjournalVolNumPages{Geophysical Research Letters}{42}{13}{5626--5634}.
\newblock
\begin{APACrefURL} \url{http://doi.wiley.com/10.1002/2015GL064260}
  \end{APACrefURL}
\newblock
\begin{APACrefDOI} \doi{10.1002/2015GL064260} \end{APACrefDOI}
\PrintBackRefs{\CurrentBib}

\bibitem [\protect \citeauthoryear {%
Satoh%
\ \protect \BOthers {.}}{%
Satoh%
\ \protect \BOthers {.}}{%
{\protect \APACyear {2019}}%
}]{%
Satoh2019}
\APACinsertmetastar {%
Satoh2019}%
\begin{APACrefauthors}%
Satoh, M.%
, Stevens, B.%
, Judt, F.%
, Khairoutdinov, M.%
, Lin, S\BHBI J.%
, Putman, W\BPBI M.%
\BCBL {}\ \BBA {} D{\"{u}}ben, P.%
\end{APACrefauthors}%
\unskip\
\newblock
\APACrefYearMonthDay{2019}{may}{}.
\newblock
{\BBOQ}\APACrefatitle {{Global Cloud-Resolving Models}} {{Global
  Cloud-Resolving Models}}.{\BBCQ}
\newblock
\APACjournalVolNumPages{Current Climate Change Reports}{}{}{1--13}.
\newblock
\begin{APACrefURL} \url{http://link.springer.com/10.1007/s40641-019-00131-0}
  \end{APACrefURL}
\newblock
\begin{APACrefDOI} \doi{10.1007/s40641-019-00131-0} \end{APACrefDOI}
\PrintBackRefs{\CurrentBib}

\bibitem [\protect \citeauthoryear {%
Selz%
, Bierdel%
\BCBL {}\ \BBA {} Craig%
}{%
Selz%
\ \protect \BOthers {.}}{%
{\protect \APACyear {2018}}%
}]{%
Selz2018}
\APACinsertmetastar {%
Selz2018}%
\begin{APACrefauthors}%
Selz, T.%
, Bierdel, L.%
\BCBL {}\ \BBA {} Craig, G\BPBI C.%
\end{APACrefauthors}%
\unskip\
\newblock
\APACrefYearMonthDay{2018}{feb}{}.
\newblock
{\BBOQ}\APACrefatitle {{Estimation of the Variability of Mesoscale Energy
  Spectra with Three Years of COSMO-DE Analyses}} {{Estimation of the
  Variability of Mesoscale Energy Spectra with Three Years of COSMO-DE
  Analyses}}.{\BBCQ}
\newblock
\APACjournalVolNumPages{Journal of the Atmospheric Sciences}{76}{2}{627--637}.
\newblock
\begin{APACrefURL}
  \url{http://journals.ametsoc.org/doi/10.1175/JAS-D-18-0155.1}
  \end{APACrefURL}
\newblock
\begin{APACrefDOI} \doi{10.1175/jas-d-18-0155.1} \end{APACrefDOI}
\PrintBackRefs{\CurrentBib}

\bibitem [\protect \citeauthoryear {%
Stein%
, Holloway%
, Tobin%
\BCBL {}\ \BBA {} Bony%
}{%
Stein%
\ \protect \BOthers {.}}{%
{\protect \APACyear {2017}}%
}]{%
Stein2017}
\APACinsertmetastar {%
Stein2017}%
\begin{APACrefauthors}%
Stein, T\BPBI H.%
, Holloway, C\BPBI E.%
, Tobin, I.%
\BCBL {}\ \BBA {} Bony, S.%
\end{APACrefauthors}%
\unskip\
\newblock
\APACrefYearMonthDay{2017}{mar}{}.
\newblock
{\BBOQ}\APACrefatitle {{Observed relationships between cloud vertical structure
  and convective aggregation over tropical ocean}} {{Observed relationships
  between cloud vertical structure and convective aggregation over tropical
  ocean}}.{\BBCQ}
\newblock
\APACjournalVolNumPages{Journal of Climate}{30}{6}{2187--2207}.
\newblock
\begin{APACrefURL}
  \url{http://journals.ametsoc.org/doi/10.1175/JCLI-D-16-0125.1}
  \end{APACrefURL}
\newblock
\begin{APACrefDOI} \doi{10.1175/JCLI-D-16-0125.1} \end{APACrefDOI}
\PrintBackRefs{\CurrentBib}

\bibitem [\protect \citeauthoryear {%
Tobin%
, Bony%
\BCBL {}\ \BBA {} Roca%
}{%
Tobin%
\ \protect \BOthers {.}}{%
{\protect \APACyear {2012}}%
}]{%
Tobin2012a}
\APACinsertmetastar {%
Tobin2012a}%
\begin{APACrefauthors}%
Tobin, I.%
, Bony, S.%
\BCBL {}\ \BBA {} Roca, R.%
\end{APACrefauthors}%
\unskip\
\newblock
\APACrefYearMonthDay{2012}{oct}{}.
\newblock
{\BBOQ}\APACrefatitle {{Observational Evidence for Relationships between the
  Degree of Aggregation of Deep Convection, Water Vapor, Surface Fluxes, and
  Radiation}} {{Observational Evidence for Relationships between the Degree of
  Aggregation of Deep Convection, Water Vapor, Surface Fluxes, and
  Radiation}}.{\BBCQ}
\newblock
\APACjournalVolNumPages{Journal of Climate}{25}{20}{6885--6904}.
\newblock
\begin{APACrefURL}
  \url{http://journals.ametsoc.org/doi/abs/10.1175/JCLI-D-11-00258.1}
  \end{APACrefURL}
\newblock
\begin{APACrefDOI} \doi{10.1175/JCLI-D-11-00258.1} \end{APACrefDOI}
\PrintBackRefs{\CurrentBib}

\bibitem [\protect \citeauthoryear {%
Torrence%
\ \BBA {} Compo%
}{%
Torrence%
\ \BBA {} Compo%
}{%
{\protect \APACyear {1998}}%
}]{%
Torrence1998}
\APACinsertmetastar {%
Torrence1998}%
\begin{APACrefauthors}%
Torrence, C.%
\BCBT {}\ \BBA {} Compo, G\BPBI P.%
\end{APACrefauthors}%
\unskip\
\newblock
\APACrefYearMonthDay{1998}{jan}{}.
\newblock
{\BBOQ}\APACrefatitle {{A Practical Guide to Wavelet Analysis}} {{A Practical
  Guide to Wavelet Analysis}}.{\BBCQ}
\newblock
\APACjournalVolNumPages{Bulletin of the American Meteorological
  Society}{79}{1}{61--78}.
\newblock
\begin{APACrefURL}
  \url{http://journals.ametsoc.org/doi/abs/10.1175/1520-0477{\%}281998{\%}29079{\%}3C0061{\%}3AAPGTWA{\%}3E2.0.CO{\%}3B2}
  \end{APACrefURL}
\newblock
\begin{APACrefDOI} \doi{10.1175/1520-0477(1998)079<0061:APGTWA>2.0.CO;2}
  \end{APACrefDOI}
\PrintBackRefs{\CurrentBib}

\bibitem [\protect \citeauthoryear {%
Trenberth%
, Stepaniak%
\BCBL {}\ \BBA {} Caron%
}{%
Trenberth%
\ \protect \BOthers {.}}{%
{\protect \APACyear {2002}}%
}]{%
Trenberth2002}
\APACinsertmetastar {%
Trenberth2002}%
\begin{APACrefauthors}%
Trenberth, K\BPBI E.%
, Stepaniak, D\BPBI P.%
\BCBL {}\ \BBA {} Caron, J\BPBI M.%
\end{APACrefauthors}%
\unskip\
\newblock
\APACrefYearMonthDay{2002}{apr}{}.
\newblock
{\BBOQ}\APACrefatitle {{Interannual variations in the atmospheric heat budget}}
  {{Interannual variations in the atmospheric heat budget}}.{\BBCQ}
\newblock
\APACjournalVolNumPages{Journal of Geophysical Research}{107}{D8}{4066}.
\newblock
\begin{APACrefURL} \url{http://doi.wiley.com/10.1029/2000JD000297}
  \end{APACrefURL}
\newblock
\begin{APACrefDOI} \doi{10.1029/2000JD000297} \end{APACrefDOI}
\PrintBackRefs{\CurrentBib}

\bibitem [\protect \citeauthoryear {%
Webster%
\ \protect \BOthers {.}}{%
Webster%
\ \protect \BOthers {.}}{%
{\protect \APACyear {1998}}%
}]{%
Webster1998}
\APACinsertmetastar {%
Webster1998}%
\begin{APACrefauthors}%
Webster, P\BPBI J.%
, Maga{\~{n}}a, V\BPBI O.%
, Palmer, T\BPBI N.%
, Shukla, J.%
, Tomas, R\BPBI A.%
, Yanai, M.%
\BCBL {}\ \BBA {} Yasunari, T.%
\end{APACrefauthors}%
\unskip\
\newblock
\APACrefYearMonthDay{1998}{jun}{}.
\newblock
{\BBOQ}\APACrefatitle {{Monsoons: Processes, predictability, and the prospects
  for prediction}} {{Monsoons: Processes, predictability, and the prospects for
  prediction}}.{\BBCQ}
\newblock
\APACjournalVolNumPages{Journal of Geophysical Research:
  Oceans}{103}{C7}{14451--14510}.
\newblock
\begin{APACrefURL} \url{http://doi.wiley.com/10.1029/97JC02719}
  \end{APACrefURL}
\newblock
\begin{APACrefDOI} \doi{10.1029/97JC02719} \end{APACrefDOI}
\PrintBackRefs{\CurrentBib}

\bibitem [\protect \citeauthoryear {%
Wheeler%
\ \BBA {} Kiladis%
}{%
Wheeler%
\ \BBA {} Kiladis%
}{%
{\protect \APACyear {1999}}%
}]{%
Wheeler1999}
\APACinsertmetastar {%
Wheeler1999}%
\begin{APACrefauthors}%
Wheeler, M.%
\BCBT {}\ \BBA {} Kiladis, G\BPBI N.%
\end{APACrefauthors}%
\unskip\
\newblock
\APACrefYearMonthDay{1999}{}{}.
\newblock
{\BBOQ}\APACrefatitle {{Convectively Coupled Equatorial Waves: Analysis of
  Clouds and Temperature in the Wavenumber-Frequency Domain}} {{Convectively
  Coupled Equatorial Waves: Analysis of Clouds and Temperature in the
  Wavenumber-Frequency Domain}}.{\BBCQ}
\newblock
\APACjournalVolNumPages{Journal of the Atmospheric Sciences}{56}{}{374--399}.
\newblock
\begin{APACrefDOI} \doi{10.1175/1520-0469(1999)056<0374:CCEWAO>2.0.CO;2}
  \end{APACrefDOI}
\PrintBackRefs{\CurrentBib}

\bibitem [\protect \citeauthoryear {%
Wielicki%
\ \protect \BOthers {.}}{%
Wielicki%
\ \protect \BOthers {.}}{%
{\protect \APACyear {1996}}%
}]{%
Wielicki1996}
\APACinsertmetastar {%
Wielicki1996}%
\begin{APACrefauthors}%
Wielicki, B\BPBI A.%
, Barkstrom, B\BPBI R.%
, Harrison, E\BPBI F.%
, Lee, R\BPBI B.%
, {Louis Smith}, G.%
\BCBL {}\ \BBA {} Cooper, J\BPBI E.%
\end{APACrefauthors}%
\unskip\
\newblock
\APACrefYearMonthDay{1996}{may}{}.
\newblock
{\BBOQ}\APACrefatitle {{Clouds and the Earth's Radiant Energy System (CERES):
  An Earth Observing System Experiment}} {{Clouds and the Earth's Radiant
  Energy System (CERES): An Earth Observing System Experiment}}.{\BBCQ}
\newblock
\APACjournalVolNumPages{Bulletin of the American Meteorological
  Society}{77}{5}{853--868}.
\newblock
\begin{APACrefURL}
  \url{http://journals.ametsoc.org/doi/abs/10.1175/1520-0477{\%}281996{\%}29077{\%}3C0853{\%}3ACATERE{\%}3E2.0.CO{\%}3B2}
  \end{APACrefURL}
\newblock
\begin{APACrefDOI} \doi{10.1175/1520-0477(1996)077<0853:CATERE>2.0.CO;2}
  \end{APACrefDOI}
\PrintBackRefs{\CurrentBib}

\bibitem [\protect \citeauthoryear {%
Wing%
, Camargo%
\BCBL {}\ \BBA {} Sobel%
}{%
Wing%
\ \protect \BOthers {.}}{%
{\protect \APACyear {2016}}%
}]{%
Wing2016}
\APACinsertmetastar {%
Wing2016}%
\begin{APACrefauthors}%
Wing, A.%
, Camargo, S\BPBI J.%
\BCBL {}\ \BBA {} Sobel, A\BPBI H.%
\end{APACrefauthors}%
\unskip\
\newblock
\APACrefYearMonthDay{2016}{jul}{}.
\newblock
{\BBOQ}\APACrefatitle {{Role of radiative-convective feedbacks in spontaneous
  tropical cyclogenesis in idealized numerical simulations}} {{Role of
  radiative-convective feedbacks in spontaneous tropical cyclogenesis in
  idealized numerical simulations}}.{\BBCQ}
\newblock
\APACjournalVolNumPages{Journal of the Atmospheric
  Sciences}{73}{7}{JAS--D--15--0380.1}.
\newblock
\begin{APACrefURL}
  \url{http://journals.ametsoc.org/doi/10.1175/JAS-D-15-0380.1}
  \end{APACrefURL}
\newblock
\begin{APACrefDOI} \doi{10.1175/JAS-D-15-0380.1} \end{APACrefDOI}
\PrintBackRefs{\CurrentBib}

\bibitem [\protect \citeauthoryear {%
Wing%
\ \BBA {} Cronin%
}{%
Wing%
\ \BBA {} Cronin%
}{%
{\protect \APACyear {2016}}%
}]{%
Wing2016a}
\APACinsertmetastar {%
Wing2016a}%
\begin{APACrefauthors}%
Wing, A.%
\BCBT {}\ \BBA {} Cronin, T\BPBI W.%
\end{APACrefauthors}%
\unskip\
\newblock
\APACrefYearMonthDay{2016}{jan}{}.
\newblock
{\BBOQ}\APACrefatitle {{Self-aggregation of convection in long channel
  geometry}} {{Self-aggregation of convection in long channel
  geometry}}.{\BBCQ}
\newblock
\APACjournalVolNumPages{Quarterly Journal of the Royal Meteorological
  Society}{142}{694}{1--15}.
\newblock
\begin{APACrefURL} \url{http://doi.wiley.com/10.1002/qj.2628} \end{APACrefURL}
\newblock
\begin{APACrefDOI} \doi{10.1002/qj.2628} \end{APACrefDOI}
\PrintBackRefs{\CurrentBib}

\bibitem [\protect \citeauthoryear {%
Wing%
, Emanuel%
, Holloway%
\BCBL {}\ \BBA {} Muller%
}{%
Wing%
\ \protect \BOthers {.}}{%
{\protect \APACyear {2017}}%
}]{%
Wing2017}
\APACinsertmetastar {%
Wing2017}%
\begin{APACrefauthors}%
Wing, A.%
, Emanuel, K.%
, Holloway, C\BPBI E.%
\BCBL {}\ \BBA {} Muller, C.%
\end{APACrefauthors}%
\unskip\
\newblock
\APACrefYearMonthDay{2017}{feb}{}.
\newblock
{\BBOQ}\APACrefatitle {{Convective Self-Aggregation in Numerical Simulations: A
  Review}} {{Convective Self-Aggregation in Numerical Simulations: A
  Review}}.{\BBCQ}
\newblock
\APACjournalVolNumPages{Surveys in Geophysics}{}{}{}.
\newblock
\begin{APACrefURL} \url{http://link.springer.com/10.1007/s10712-017-9408-4}
  \end{APACrefURL}
\newblock
\begin{APACrefDOI} \doi{10.1007/s10712-017-9408-4} \end{APACrefDOI}
\PrintBackRefs{\CurrentBib}

\bibitem [\protect \citeauthoryear {%
Wing%
\ \BBA {} Emanuel%
}{%
Wing%
\ \BBA {} Emanuel%
}{%
{\protect \APACyear {2014}}%
}]{%
Wing2014}
\APACinsertmetastar {%
Wing2014}%
\begin{APACrefauthors}%
Wing, A.%
\BCBT {}\ \BBA {} Emanuel, K\BPBI a.%
\end{APACrefauthors}%
\unskip\
\newblock
\APACrefYearMonthDay{2014}{}{}.
\newblock
{\BBOQ}\APACrefatitle {{Physical mechanisms controlling self-aggregation of
  convection in idealized numerical modeling simulations}} {{Physical
  mechanisms controlling self-aggregation of convection in idealized numerical
  modeling simulations}}.{\BBCQ}
\newblock
\APACjournalVolNumPages{Journal of Advances in Modeling Earth
  Systems}{5}{November}{59--74}.
\newblock
\begin{APACrefURL} \url{http://dx.doi.org/10.1002/2013MS000269}
  \end{APACrefURL}
\newblock
\begin{APACrefDOI} \doi{10.1002/2013MS000269} \end{APACrefDOI}
\PrintBackRefs{\CurrentBib}

\bibitem [\protect \citeauthoryear {%
Wing%
\ \protect \BOthers {.}}{%
Wing%
\ \protect \BOthers {.}}{%
{\protect \APACyear {2018}}%
}]{%
Wing2018}
\APACinsertmetastar {%
Wing2018}%
\begin{APACrefauthors}%
Wing, A.%
, Reed, K\BPBI A.%
, Satoh, M.%
, Stevens, B.%
, Bony, S.%
\BCBL {}\ \BBA {} Ohno, T.%
\end{APACrefauthors}%
\unskip\
\newblock
\APACrefYearMonthDay{2018}{}{}.
\newblock
{\BBOQ}\APACrefatitle {{Radiative-convective equilibrium model intercomparison
  project}} {{Radiative-convective equilibrium model intercomparison
  project}}.{\BBCQ}
\newblock
\APACjournalVolNumPages{Geoscientific Model Development}{11}{2}{793--813}.
\newblock
\begin{APACrefURL} \url{https://doi.org/10.5194/gmd-11-793-2018}
  \end{APACrefURL}
\newblock
\begin{APACrefDOI} \doi{10.5194/gmd-11-793-2018} \end{APACrefDOI}
\PrintBackRefs{\CurrentBib}

\bibitem [\protect \citeauthoryear {%
Woodhams%
\ \protect \BOthers {.}}{%
Woodhams%
\ \protect \BOthers {.}}{%
{\protect \APACyear {2018}}%
}]{%
Woodhams2018}
\APACinsertmetastar {%
Woodhams2018}%
\begin{APACrefauthors}%
Woodhams, B\BPBI J.%
, Birch, C\BPBI E.%
, Marsham, J\BPBI H.%
, Bain, C\BPBI L.%
, Roberts, N\BPBI M.%
\BCBL {}\ \BBA {} Boyd, D\BPBI F\BPBI A.%
\end{APACrefauthors}%
\unskip\
\newblock
\APACrefYearMonthDay{2018}{sep}{}.
\newblock
{\BBOQ}\APACrefatitle {{What Is the Added Value of a Convection-Permitting
  Model for Forecasting Extreme Rainfall over Tropical East Africa?}} {{What Is
  the Added Value of a Convection-Permitting Model for Forecasting Extreme
  Rainfall over Tropical East Africa?}}{\BBCQ}
\newblock
\APACjournalVolNumPages{Monthly Weather Review}{146}{9}{2757--2780}.
\newblock
\begin{APACrefURL}
  \url{http://journals.ametsoc.org/doi/10.1175/MWR-D-17-0396.1}
  \end{APACrefURL}
\newblock
\begin{APACrefDOI} \doi{10.1175/mwr-d-17-0396.1} \end{APACrefDOI}
\PrintBackRefs{\CurrentBib}

\bibitem [\protect \citeauthoryear {%
Yasunaga%
, Yokoi%
, Inoue%
\BCBL {}\ \BBA {} Mapes%
}{%
Yasunaga%
\ \protect \BOthers {.}}{%
{\protect \APACyear {2019}}%
}]{%
Yasunaga2019}
\APACinsertmetastar {%
Yasunaga2019}%
\begin{APACrefauthors}%
Yasunaga, K.%
, Yokoi, S.%
, Inoue, K.%
\BCBL {}\ \BBA {} Mapes, B\BPBI E.%
\end{APACrefauthors}%
\unskip\
\newblock
\APACrefYearMonthDay{2019}{jan}{}.
\newblock
{\BBOQ}\APACrefatitle {{Space-time spectral analysis of the moist static energy
  budget equation}} {{Space-time spectral analysis of the moist static energy
  budget equation}}.{\BBCQ}
\newblock
\APACjournalVolNumPages{Journal of Climate}{32}{2}{501--529}.
\newblock
\begin{APACrefURL}
  \url{http://journals.ametsoc.org/doi/10.1175/JCLI-D-18-0334.1}
  \end{APACrefURL}
\newblock
\begin{APACrefDOI} \doi{10.1175/JCLI-D-18-0334.1} \end{APACrefDOI}
\PrintBackRefs{\CurrentBib}

\bibitem [\protect \citeauthoryear {%
Zhang%
}{%
Zhang%
}{%
{\protect \APACyear {2005}}%
}]{%
Zhang2005}
\APACinsertmetastar {%
Zhang2005}%
\begin{APACrefauthors}%
Zhang, C.%
\end{APACrefauthors}%
\unskip\
\newblock
\APACrefYearMonthDay{2005}{}{}.
\newblock
{\BBOQ}\APACrefatitle {{Madden-Julian Oscillation}} {{Madden-Julian
  Oscillation}}.{\BBCQ}
\newblock
\APACjournalVolNumPages{Reviews of Geopyhsics}{43}{2004}{1--36}.
\newblock
\begin{APACrefDOI} \doi{10.1029/2004RG000158.1.INTRODUCTION} \end{APACrefDOI}
\PrintBackRefs{\CurrentBib}

\end{thebibliography}
%




\end{document}